\documentclass[10pt,twocolumn,letterpaper]{article}

\usepackage{cvpr}              

\makeatletter
\newif\if@restonecol
\makeatother

\usepackage[linesnumbered,ruled, vlined]{algorithm2e}
\usepackage{algpseudocode}
\usepackage{amsmath}

\usepackage{amsmath}
\usepackage{amssymb}
\usepackage{booktabs}
\usepackage{stfloats}
\usepackage{multirow}
\usepackage{graphicx}
\usepackage{float}
\usepackage{threeparttable}

\usepackage{marvosym}
\usepackage{ifsym}
\usepackage[accsupp]{axessibility}  

\usepackage[pagebackref,breaklinks,colorlinks]{hyperref}
\usepackage[labeled]{multibib}
\newcites{A}{Appendix References}

\usepackage[capitalize]{cleveref}
\crefname{section}{Sec.}{Secs.}
\Crefname{section}{Section}{Sections}
\Crefname{table}{Table}{Tables}
\crefname{table}{Tab.}{Tabs.}

\def\cvprPaperID{7523} 
\def\confName{CVPR}
\def\confYear{2023}

\begin{document}

\title{QPGesture: Quantization-Based and Phase-Guided Motion Matching for Natural Speech-Driven Gesture Generation}     

\author{Sicheng Yang$^{1}$ \quad \text{Zhiyong Wu}$^{\textrm{\Letter},1,4}$ \quad Minglei Li$^{2}$ \quad Zhensong Zhang$^{3}$ \\
Lei Hao$^{3}$ \quad Weihong Bao$^{1}$ \quad Haolin Zhuang$^{1}$
\\      
$^{1} $ Tsinghua Shenzhen International Graduate School, Tsinghua University \\       
$^{2} $ Huawei Cloud Computing Technologies Co., Ltd
$^{3} $ Huawei Noah’s Ark Lab \\
$^{4} $ The Chinese University of Hong Kong\\
{\tt\small \{yangsc21, bwh21, zhuanghl21\}@mails.tsinghua.edu.cn}
\quad {\tt\small zywu@sz.tsinghua.edu.cn} \\
{\tt\small \{liminglei29, zhangzhensong, haolei5\}@huawei.com}
}


\maketitle

\begin{abstract}
    Speech-driven gesture generation is highly challenging due to the random jitters of human motion.
   In addition, there is an inherent asynchronous relationship between human speech and gestures.
   To tackle these challenges, we introduce a novel quantization-based and phase-guided motion matching framework. Specifically, we first present a gesture VQ-VAE module to learn a codebook to summarize meaningful gesture units. With each code representing a unique gesture, random jittering problems are alleviated effectively. We then use Levenshtein distance to align diverse gestures with different speech. Levenshtein distance based on audio quantization as a similarity metric of corresponding speech of gestures helps match more appropriate gestures with speech, and solves the alignment problem of speech and gestures well. Moreover, we introduce phase to guide the optimal gesture matching based on the semantics of context or rhythm of audio. Phase guides when text-based or speech-based gestures should be performed to make the generated gestures more natural. Extensive experiments show that our method outperforms recent approaches on speech-driven gesture generation.
   Our code, database, pre-trained models and demos are available at \url{https://github.com/YoungSeng/QPGesture}.
   
\end{abstract}

\begin{figure}[!t]
  \centering
   \includegraphics[width=\linewidth]{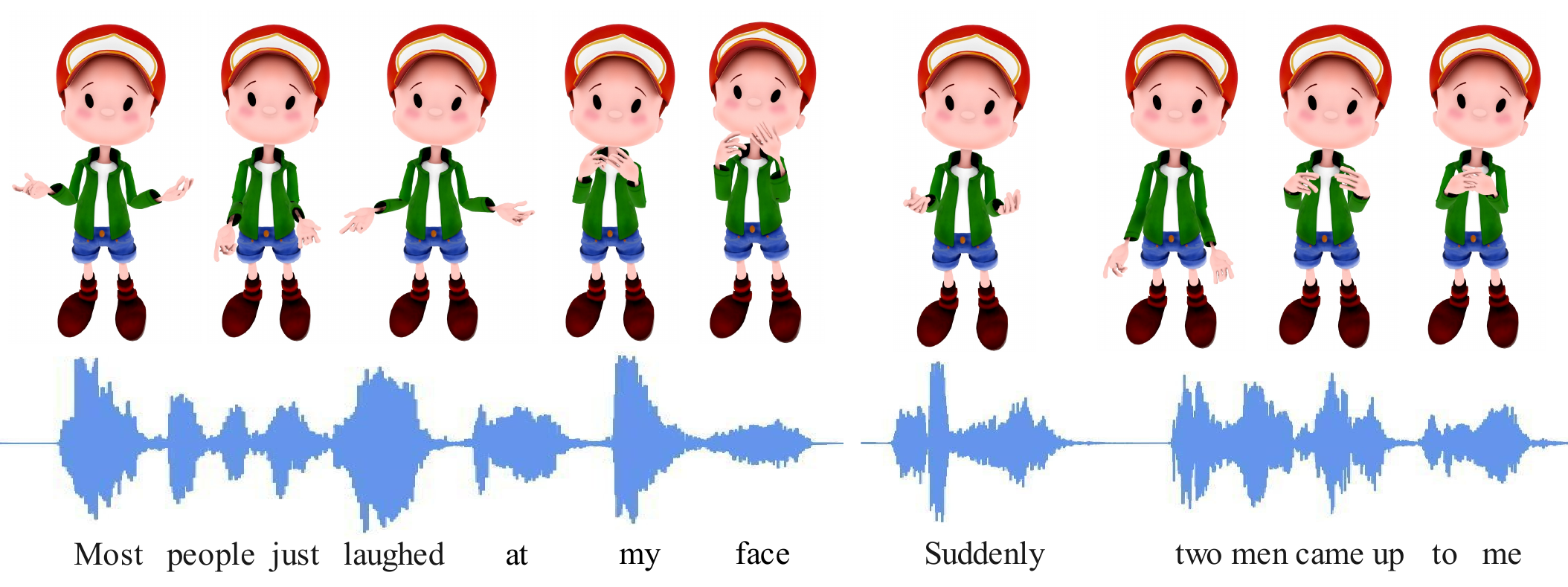}
   \caption{\textbf{Gesture examples generated by our proposed method} on various types of speech. 
  The character is from Mixamo \cite{Mixamo}.
}
   \label{fig:overview}
\end{figure}

\section{Introduction}

Nonverbal behavior plays a key role in conveying messages in human communication \cite{DBLP:conf/iui/KucherenkoJYWH21}, including facial expressions, hand gestures and body gestures.
Co-speech gesture helps better self-expression \cite{DBLP:conf/icmi/Yang0LZLCB22}.
However, producing human-like and speech-appropriate gestures is still very difficult due to two main challenges: 
\textbf{1) Random jittering}: People make many small jitters and movements when they speak, which can lead to a decrease in the quality of the generated gestures.
\textbf{2) Inherent asynchronicity with speech}: Unlike speech with face or lips, there is an inherent asynchronous relationship between human speech and gestures.

Most existing gesture generation studies intend to solve the two challenges in a single ingeniously designed neural network that directly maps speech to 3D joint sequence in high-dimensional continuous space \cite{DBLP:conf/atal/KucherenkoNNKH22, DBLP:journals/ijhci/KucherenkoHKHK21, DBLP:journals/corr/abs-2203-05297, DBLP:conf/iva/HabibieXMLSPET21} using a sliding window with a fixed step size \cite{Yoon2020Speech, DBLP:conf/icra/YoonKJLKL19, DBLP:conf/siggraph/HabibieESANNT22}.
However, such methods are limited by the representation power of proposed neural networks, like the GENEA gesture-generation challenge results.
No system in GENEA challenge 2020 \cite{DBLP:conf/iui/KucherenkoJYWH21} rated above a bottom line that paired the input speech audio with mismatched excerpts of training data motion.
In GENEA challenge 2022 \cite{DBLP:conf/icmi/YoonWKVNTH22}, a motion matching based method \cite{DBLP:conf/icmi/ZhouBC22} ranked first in the human-likeness evaluation and upper-body appropriateness evaluation, and outperformed all other neural network-based models.
These results indicate that motion matching based models, if designed properly, are more effective than neural network based models.

Inspired by this observation, in this work, we propose a novel quantization-based motion matching framework for audio-driven gesture generation. Our framework includes two main components aiming at solving the two above challenges, respectively.
First, to address the random jittering challenge, we compress human gestures into a space that is lower dimensional and discrete, to reduce input redundancy. Instead of manually indicating the gesture units \cite{DBLP:phd/de/Kipp2007}, we use a vector quantized variational autoencoder (VQ-VAE) \cite{van2017neural} to encode and quantize joint sequences to a codebook in an unsupervised manner, using a quantization bottleneck.
Each learned code is shown to represent a unique gesture pose.
By reconstructing the discrete gestures, some random jittering problems such as grabbing hands and pushing glasses will be solved.
Second, to address the inherent asynchronicity of speech and gestures, Levenshtein distance \cite{levenshtein1966binary} is used based on audio quantization.
Levenshtein distance helps match more appropriate gestures with speech, and solves the alignment problem of speech and gestures well.
Moreover, unlike the recent gesture matching models \cite{DBLP:conf/siggraph/HabibieESANNT22, DBLP:conf/icmi/ZhouBC22}, we also consider the semantic information of the context. 
Third,
since the body motion is composed of multiple periodic motions spatially, meanwhile the phase values are able to describe the nonlinear periodicity of the high-dimensional motion curves well \cite{10.1145/3528223.3530178},
we use phase to guide how the gestures should be matched to speech and text.

\begin{figure*}[!t]
  \centering
   \includegraphics[width=\linewidth]{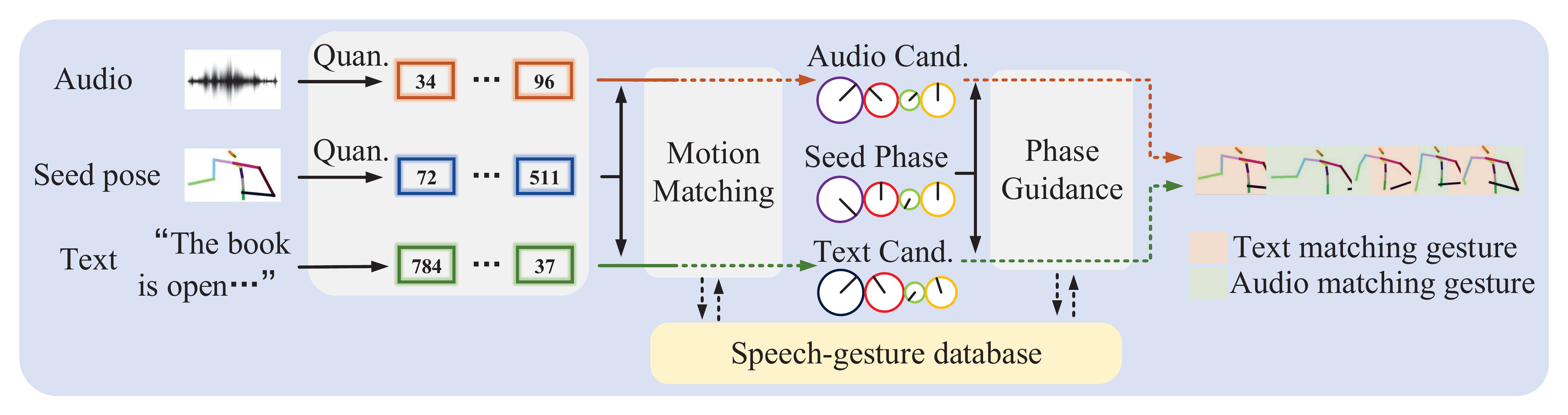}
   \caption{\textbf{Gesture generation pipeline of our proposed framework.}
   `Quan.' is short for `quantization' and `Cand.' is short for `candidate'.
   Given a piece of audio, text and seed pose, the audio and gesture are quantized. 
   The candidate for the speech is calculated based on the Levenshtein distance, and the candidate for the text is calculated based on the cosine similarity. 
   The optimal gesture is selected based on phase-guidance corresponding to the seed code and the phase corresponding to the two candidates.
}
   \label{fig:pipeline}
\end{figure*}

The inference procedure of our framework is shown in Figure \ref{fig:pipeline}.
Given a piece of audio, text and seed pose, the audio and gesture are first quantized.
The best candidate for the speech is calculated based on the Levenshtein distance, and the best candidate for the text is calculated based on the cosine similarity. 
Then the most optimal gesture is selected based on the phase corresponding to the seed code and the phase corresponding to the two candidates.

The main contributions of our work are:
\begin{itemize}
	\item We present a novel quantization-based motion matching framework for speech-driven gesture generation. 
    \item We propose to align diverse gestures with different speech using Levenshtein distance, based on audio quantization.
    \item We design a phase guidance strategy to select optimal audio and text candidates for motion matching.
    \item Extensive experiments show that jittering and asynchronicity issues can be effectively alleviated by our framework.
\end{itemize}

\section{Related Work}

\textbf{End-to-end Co-speech Gesture Generation.}
Gesture generation is a complex problem that requires understanding speech, gestures, and their relationships.
Data-driven approaches attempt to learn gesticulation skills from human demonstrations.
The present studies mainly consider four modalities: text \cite{DBLP:journals/corr/abs-2101-05684, DBLP:conf/icra/YoonKJLKL19, DBLP:conf/icmi/WangAGBHS21}, audio \cite{DBLP:conf/iccv/QianTZ0G21, DBLP:conf/iva/HabibieXMLSPET21, DBLP:journals/ijhci/KucherenkoHKHK21, DBLP:conf/iccv/0071KPZZ0B21, DBLP:conf/cvpr/GinosarBKCOM19}, gesture motion, and speaker identity\cite{Yoon2020Speech, DBLP:journals/corr/abs-2203-05297, DBLP:conf/cvpr/LiuWZXQLZWDZ22, DBLP:conf/eccv/AhujaLNM20, DBLP:journals/cgf/AlexandersonHKB20, DBLP:journals/corr/abs-2203-02291, DBLP:conf/mm/BhattacharyaCRM21}. 
Habibie \etal \cite{DBLP:conf/iva/HabibieXMLSPET21} propose the first approach to jointly synthesize both the synchronous 3D conversational body and hand gestures, as well as 3D face and head animations.
Ginosar \etal \cite{DBLP:conf/cvpr/GinosarBKCOM19} propose a cross-modal translation method based on the speech-driven gesture gestures of a single speaker.
Liu \etal \cite{DBLP:conf/cvpr/LiuWZXQLZWDZ22} propose a hierarchical audio learner extracts audio representations across semantic granularities and a hierarchical pose inferior renders the entire human pose. 
Kucherenko \etal \cite{DBLP:journals/ijhci/KucherenkoHKHK21} propose Aud2Repr2Pose architecture to evaluate the impact of different gesture and speech representations on gesture generation.
Qian \etal \cite{DBLP:conf/iccv/QianTZ0G21} use conditional learning to resolve the ambiguity of co-speech gesture synthesis by learning the template vector to improve gesture quality.

As for learning individual styles, 
Yoon \etal \cite{Yoon2020Speech} propose the first end-to-end method for generating co-speech gestures using the tri-modality of text, audio and speaker identity.
Ahuja \etal \cite{DBLP:conf/eccv/AhujaLNM20} train a single model for multiple speakers while learning style embeddings for gestures of each speaker.
Alexanderson \etal \cite{DBLP:journals/cgf/AlexandersonHKB20} adapt MoGlow to speech-driven gesture
synthesis and added a framework for high-level control the gesturing style.
Liang \etal \cite{DBLP:conf/cvpr/LiangFZHP022} propose a semantic energized generation method for semantic-aware gesture generation.
Li \etal \cite{DBLP:conf/iccv/0071KPZZ0B21} propose a conditional variational autoencoder that models one-to-many audio-to-motion mapping by splitting the cross-modal latent code into shared code and motion-specific code.

\begin{figure}[t]
  \centering
   \includegraphics[width=0.99\linewidth]{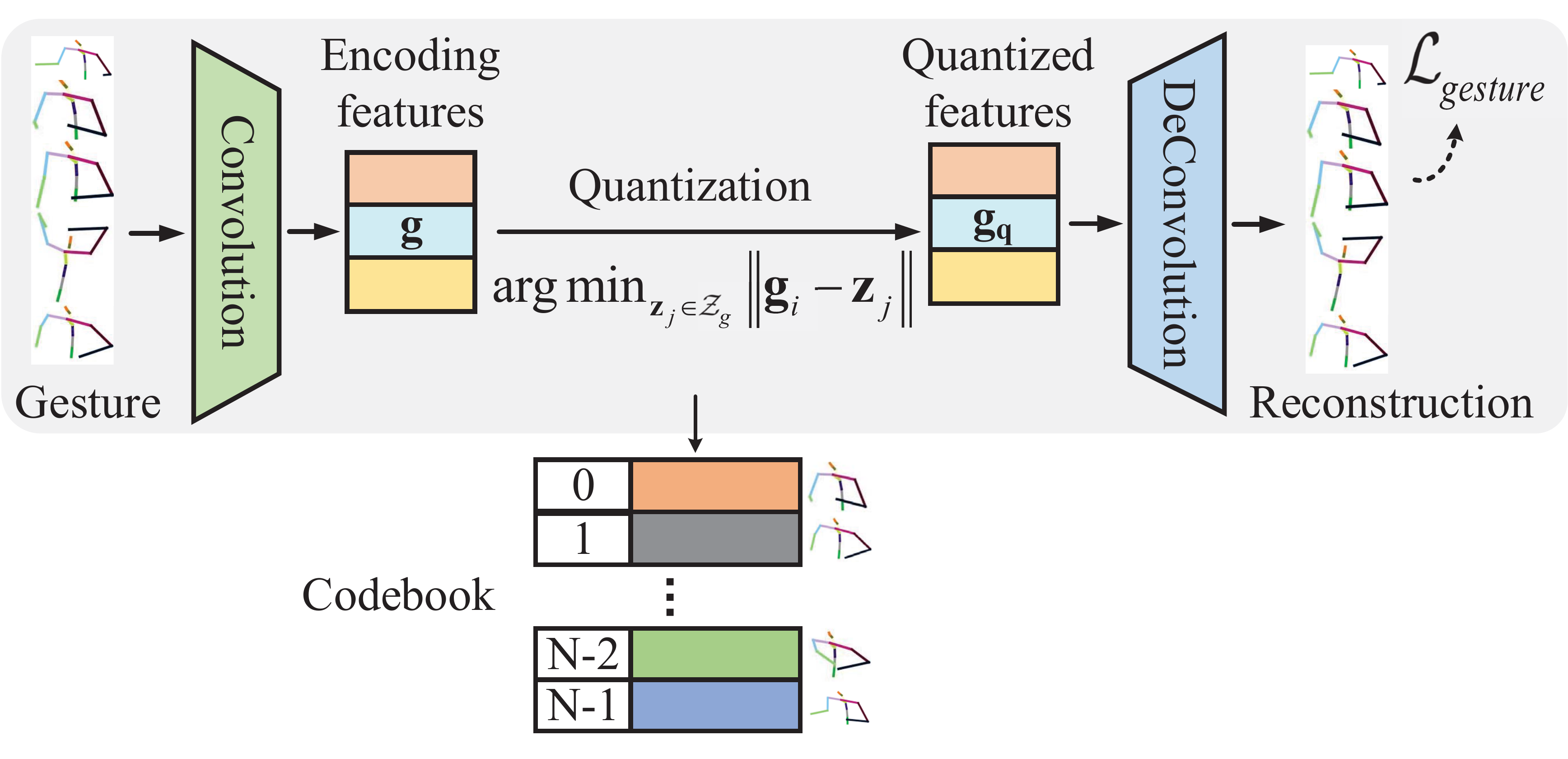}
   \caption{\textbf{Structure of gesture VQ-VAE.} After learning the discrete latent representation of human gesture, the gesture VQ-VAE encode and summarize meaningful gesture units, and reconstruct the target gesture sequence from quantized latent features.}
   \label{fig:motionvq}
\end{figure}

Michael \etal \cite{buttner2015motion} propose \textit{motion matching}, which is a k-Nearest Neighbor (KNN) search method for searching a large database of animations.
Zhou \etal \cite{DBLP:conf/cvpr/ZhouYLSAK22} utilize a graph-based framework to synthesize body motions for conversations. 
Ferstl \etal \cite{DBLP:journals/jvca/FerstlNM21} predict expressive gestures based on database matching. 
Habibie \etal \cite{DBLP:conf/siggraph/HabibieESANNT22} predicted motion using a KNN algorithm and use a conditional generative adversarial network to refine the result.
Zhou \etal \cite{DBLP:conf/icmi/ZhouBC22} calculate rhythm signature and style signature using StyleGestures \cite{DBLP:journals/cgf/AlexandersonHKB20}, and synthesized graph-based matching gesture.
Recently, a large-scale 3D gestures dataset BEAT \cite{DBLP:journals/corr/abs-2203-05297} is built from multi-camera videos based on six modalities of data, which we used in this task.

\textbf{Quantization-based Pose representation.}
Kipp has represented gestures as predefined unit gestures \cite{DBLP:phd/de/Kipp2007}.
Van \etal \cite{van2017neural} propose Vector VQ-VAE to generate discrete representations.
Guo \etal \cite{DBLP:journals/corr/abs-2207-01696} use motion tokens to generate human full-body motions from texts, and their reciprocal task.
Lucas \etal \cite{lucas2022posegpt} propose to train a GPT-like model for next-index prediction in that space.
Hong \etal \cite{DBLP:journals/tog/HongZPCYL22} use a pose codebook created by clustering to generate diverse poses.
Li \etal \cite{siyao2022bailando} propose to pose VQ-VAE to encode and summarize dancing units.
Existing studies have shown that quantification helps to reduce motion freezing during motion generation and retains the details of motion well \cite{DBLP:journals/corr/abs-2210-01448}.

In our work, we encode and quantize meaningful gesture constituents and generate human-like gestures by speech based motion matching with phase guidance.

\section{Our Approach}
Our approach takes a piece of audio, corresponding text, seed pose and optionally a sequence of control signals as inputs, and outputs a sequence of gesture motion. We first prepossess these inputs as well as the motion database into their quantized discrete forms automatically. We then find the best candidate for the speech and the best candidate for the text, respectively. Finally, we select the optimal gestures based on the phase corresponding to the seed code and the phase corresponding to the two candidates. The rest of this section describes the details of each step.
\subsection{Learning a discrete latent space representation}

\textbf{Gesture Quantization.}
We design a pose VQ-VAE as shown in Figure \ref{fig:motionvq}.
Given the gesture sequence $\mathbf{G} \in \mathbb{R}^{T \times D_g}$, where $T$ denotes the number of gestures and $D_g$ pose dimension. 
We first adopt a 1D temporal convolution network $E_g$ to encode the joint sequence $\mathbf{G}$ to context-aware features $\mathbf{g} \in \mathbb{R}^{T^{\prime} \times C}$, where $T^{\prime}=T / d$, $d$ is the temporal down-sampling rate, and $C$ is the channel dimension of features.
This process could be written as $\mathbf{g}=E_g(\mathbf{G})$. 
Codebook could index an embedding table with samples drawn from the distributions \cite{van2017neural}.
To learn the corresponding codebook $\mathcal{Z}_g$ elements $\mathbf{g}_{\mathbf{q}} \in \mathbb{R}^{T^{'} \times C}$, where $\mathbf{g}_q \in \mathcal{Z}_{g}^{T}$, $\mathcal{Z}_g$ is a set of $C_b$ codes of dimension $n_z$, we quantize $\mathbf{g}$ by mapping each temporal feature $\mathbf{g}_i$ to its closest codebook element $z_j$ as $\mathbf{q}(.)$:
\begin{equation}
\mathbf{g}_{\mathbf{q}, i}=\mathbf{q}(\mathbf{g}) = \arg \min _{\mathbf{z}_j \in \mathcal{Z}_g}\left\|\mathbf{g}_i-\mathbf{z}_j\right\|
\end{equation}

A following de-convolutional decoder $D_g$ projects $\mathbf{g}_{\mathbf{q}}$ back to the motion space as a pose sequence $\hat{\mathbf{G}_1}$, which can be formulated as
\begin{equation}
\hat{\mathbf{G}_1}=D_g\left(\mathbf{g}_q\right)=D_g(\mathbf{q}(E_g(\mathbf{G})))
\end{equation}

Thus the encoder, decoder and codebook can be trained by optimizing:
\begin{equation}
\label{beta}
\begin{split}
\mathcal{L}_{gesture(E_g, D_g, \mathcal{Z}_g)}=\mathcal{L}_{r e c}(\hat{\mathbf{G}_1}, \mathbf{G})+\left\|\operatorname{sg}[\mathbf{g}]-\mathbf{g}_{\mathbf{q}}\right\|\\+\beta\left\|\mathbf{g}-\operatorname{sg}\left[\mathbf{g}_{\mathbf{q}}\right]\right\|
\end{split}
\end{equation}
where $\mathcal{L}_{r e c}$ is the reconstruction loss that constrains the predicted joint sequence to ground truth,  $\operatorname{sg}[\cdot]$ denotes the stop-gradient operation, and the term $\left\|\mathbf{g}-\operatorname{sg}\left[\mathbf{g}_{\mathbf{q}}\right]\right\|$ is the “commitment loss” with weighting factor $\beta$ \cite{van2017neural}.

Inspired by Bailando \cite{siyao2022bailando}, a music-driven dance model, we add velocity loss and acceleration loss to the reconstruction loss to prevent jitters in generated gesture:
\begin{equation}
\label{alpha}
\begin{split}
\mathcal{L}_{r e c}(\hat{\mathbf{G}_1}, \mathbf{G}_1)=\|\hat{\mathbf{G}_1}-\mathbf{G}_1\|_1+\alpha_1\left\|\hat{\mathbf{G}_1}^{\prime}-\mathbf{G}_1^{\prime}\right\|_1\\
+\alpha_2\left\|\hat{\mathbf{G}_1}^{\prime \prime}-\mathbf{G}_1^{\prime \prime}\right\|_1
\end{split}
\end{equation}

And to avoid encoding confusion caused by the global shift of joints, we normalize the absolute locations of input $\mathbf{G}$ \ie, set the root joints (hips) to 0, and make objects face the same direction.
Standard normalization (zero mean and unit variant) is applied to all joints.

\textbf{Audio Quantization.}
We use vq-wav2vec Gumbel-Softmax model \cite{DBLP:conf/iclr/BaevskiSA20} pre-trained on a clean 100h subset of Librispeech \cite{DBLP:conf/icassp/PanayotovCPK15} which is discretized to 102.4K tokens.
We multiply the values of the two groups together as the token of the segment of audio.
The convolutional encoder produces a representation $\mathbf{z}$, for each time step $i$, 
the quantization module replaces the original representation $\mathbf{z}$ by $\hat{\mathbf{z}} = \mathbf{a}_{\mathbf{q}, i}$ from $\mathcal{Z}_{a}$, which contains a set of $C_b^{'}$ codes of dimension $n_z^{'}$.

\subsection{Motion Matching based on Audio and Text}

Our motion matching algorithm takes a discrete text sequence $\mathbf{t}=[\mathbf{t}_0, \mathbf{t}_1, \dots, \mathbf{t}_{T^{\prime}-1}]$, a discrete audio sequence $\mathbf{a}_{\mathbf{q}}=[\mathbf{a}_{\mathbf{q}, 0}, \mathbf{a}_{\mathbf{q}, 1}, \dots, \mathbf{a}_{\mathbf{q}, T^{\prime}-1}]$, and one initial previous pose code $\mathbf{g}_{-1}$, and optionally a sequence of control masks $\mathbf{M} = [\mathbf{m}_0, \mathbf{m}_1, \dots, \mathbf{m}_{T^{\prime}-1}]$.
The outputs are audio-based candidate $\mathbf{C}_a = [\mathbf{c}_{0,a}, \mathbf{c}_{1,a}, \dots, \mathbf{c}_{T^{\prime}-1,a}]$ and text-based candidate $\mathbf{C}_t = [\mathbf{c}_{0,t}, \mathbf{c}_{1,t}, \dots, \mathbf{c}_{T^{\prime}-1,t}]$. $T^{\prime}$ denotes the number of speech segments during inference.

Considering too long gesture clips time decreases the diversity of gestures and too short gesture clips time results in poor human likeness \cite{DBLP:conf/icmi/ZhouBC22}, we split each gesture motion in the dataset into clip-level gesture clips automatically by time interval of words larger than 0.5 seconds in text transcriptions. 
These clip-level data form the speech-gesture database in Figure \ref{fig:overview}.

To find the appropriate output gesture sequence $\mathbf{C}_a$ and $\mathbf{C}_t$ from the database, we consider both the similarity with respect to the current test speech as well as the previously searched pose code $\mathbf{g}_{-1}$ for every $T$ frame interval for better continuity between consecutive syntheses, as in \cite{Yoon2020Speech}.
During each iteration, we first compare the Euclidean distance between the joints corresponding to initial code $\mathbf{g}_{-1}$ and the joints corresponding to each code $\mathbf{g}_{\mathbf{q}}$ in the codebook to get the pose-based pre-candidate $\hat{\mathbf{C}}_{g} = \left\{\hat{\mathbf{C}}_{g}^0, \hat{\mathbf{C}}_{g}^1, \ldots, \hat{\mathbf{C}}_{g}^{C_b}\right\}$.
In the first iteration, the previous pose code $\mathbf{g}_{-1}$ is initialized by either randomly sampling a code from codebook $\mathcal{Z}_g$ or set to be the code with the most frequent occurrences in the codebook (the code of mean pose). 

\textbf{Audio-based Search.} 
To encode information about the relevant past and future codes, we use a search-centered window (0.5 seconds long) as a feature for the current time, then we get audio features sequence $\mathbf{F}_{a}=[\mathbf{F}_{a, 0}, \mathbf{F}_{a, 1}, \dots, F_{a, T^{\prime}-1}]$.
To address the inherent asynchronicity of speech and gestures, Levenshtein distance \cite{levenshtein1966binary} is used to measure the similarity of the test audio by comparing the current test audio feature with the audio feature of clips from the database. 
For every clip in the database, we calculate the audio feature similarity of the corresponding code per $d$ frame and take the minimum value as the audio candidate distance for each code.      
Then we get audio-based pre-candidate $\hat{\mathbf{C}}_{a}=\left\{\hat{\mathbf{C}}_{a}^0, \hat{\mathbf{C}}_{a}^1, \ldots, \hat{\mathbf{C}}_{a}^{C_b}\right\}$.
Levenshtein distance as a similarity metric of corresponding speech of gestures helps match more appropriate gestures with speech, and solves the alignment problem of speech and gestures well (see Section \ref{section:Ablation}).


\textbf{Text-based Search.} 
Similarly, we use the text before and after 0.5 seconds as the sentence of the current code. 
To obtain the semantic information of the context, we use Sentence-BERT \cite{DBLP:conf/emnlp/ReimersG19} to compute sentence embeddings, as text features sequence $\mathbf{F}_{t}$.
For every clip in the database, we calculate the text features cosine similarity of the corresponding code per $d$ frame and take the minimum value as the text candidate distance for each code. Then text-based pre-candidate is $\hat{\mathbf{C}}_{t}=\left\{\hat{\mathbf{C}}_{t}^0, \hat{\mathbf{C}}_{t}^1, \ldots, \hat{\mathbf{C}}_{t}^{C_b}\right\}$

\begin{figure}[t]
  \centering
   \includegraphics[width=0.99\linewidth]{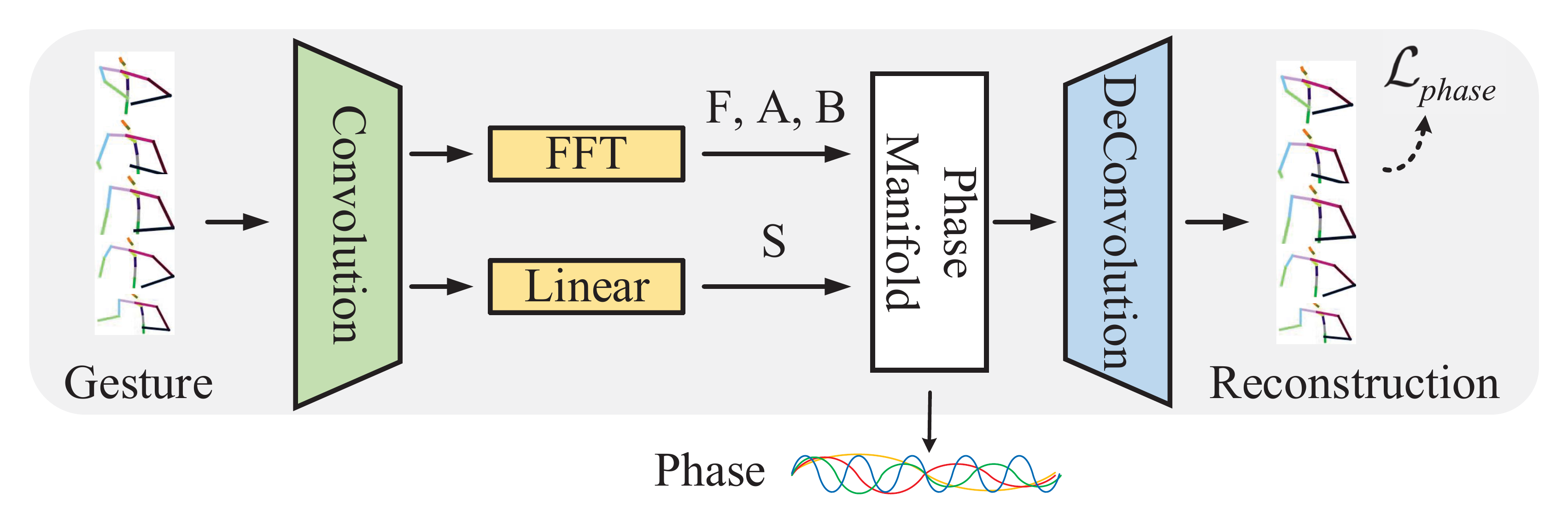}
   \caption{\textbf{Architecture of the Periodic Autoencoder Network.} The convolutional network encoder learns a lower dimensional embedding of the gesture. Then differentiable real Fast Fourier Transform (FFT) and fully connected network is applied to get periodic parameters: amplitude (A), frequency (F), offset (B) and phase shift (S). The deconvolutional network decoder map all periodic parameters back to the original motion curves to force the periodic parameters to reconstruct the original latent embedding.}
   \label{fig:deepphase}
\end{figure}

Instead of weighting the similarity score of audio/text and pose terms, we add pose similarity ranks and audio/text similarity ranks for every pre-candidate to select the lowest rank \cite{DBLP:conf/siggraph/HabibieESANNT22} result as the audio/text candidate.
The generated gesture will be discontinuous if only speech (audio and text) matching is considered, so as shown in the figure \ref{fig:pipeline}, we first compute the pre-candidates of pose, audio, and text, then compute audio candidate $\mathbf{C}_a$ based on $\hat{\mathbf{C}}_{g}$ and $\hat{\mathbf{C}}_{a}$, and text candidate $\mathbf{C}_t$ based on $\hat{\mathbf{C}}_{g}$ and $\hat{\mathbf{C}}_{t}$.
Then we select the final gesture from $\mathbf{C}_a$ and $\mathbf{C}_t$ according to the phase guide.

\subsection{Phase-Guided Gesture Generation}

To transform the motion space into a learned phase manifold, we utilize a temporal periodic autoencoder architecture structure similar to DeepPhase \cite{10.1145/3528223.3530178}.
The architecture is shown in figure \ref{fig:deepphase}.
First, we adopt a 1D temporal convolution network $E_p$ to a latent space of the motion $\mathbf{G}$, which can be formulated as 
\begin{equation}
\mathbf{L}=E_p(\mathbf{G})
\end{equation}
where $\mathbf{L} \in \mathbb{R}^{T \times M}$, $M$ is the number of latent channels, that is, the number of desired phase channels to be extracted from the motion.
For each channel, our goal is to extract a good phase offset to capture its current point as part of a larger cycle.

It is complicated to calculate the phase of a cluttered curve directly, so we calculate periodic parameters amplitude (A), frequency (F), offset (B) and phase shift (S) first.
We apply differentiable real Fast Fourier Transform (FFT) to each channel of $\mathbf{L}$ and create the zero-indexed matrix of Fourier coefficients $\mathbf{c} \in \mathbb{C}^{M \times K+1}$, $K=\left\lfloor\frac{T}{2}\right\rfloor$, written as $\mathbf{c}=F F T(\mathbf{L})$. Power spectrum $\mathbf{p} \in \mathbb{R}^{M \times K+1}$ for each channel is $\mathbf{p}_{i, j}=\frac{2}{T}\left|\mathbf{c}_{i, j}\right|^2$, where $i$ is the channel index and $j$ is the index for the frequency bands. 
Assumed that there are $T$ points in a time window of $N$ seconds.
The periodic parameters are computed by:
\begin{equation}
\mathbf{A}_i=\sqrt{\frac{2}{T} \sum_{j=1}^K \mathbf{p}_{i, j}}, \quad \mathbf{F}_i=\frac{\sum_{j=1}^K\left(\mathbf{f}_j \cdot \mathbf{p}_{i, j}\right)}{\sum_{j=1}^K \mathbf{p}_{i, j}}, \quad \mathbf{B}_i=\frac{\mathbf{c}_{i, 0}}{T},
\end{equation}
where $\mathbf{f}=(0,1 / N, 2 / N, \ldots, K / N)$ is frequencies vector.

Phase shift $\mathbf{S} \in \mathbb{R}^M$ for each latent curve $\mathbf{S}_i$ can be predicted by a fully connected layer $FC$, which can be formulated as:
\begin{equation}
\left(s_x, s_y\right)=F C\left(\mathbf{L}_i\right), \quad \mathbf{S}_i=\operatorname{atan} 2\left(s_y, s_x\right)
\end{equation}

For each temporal motion $\mathbf{G}$, within a centered time window $\mathcal{T}=\left[-\frac{t_1-t_0}{2},-\frac{t_1-t_0}{2}+\frac{t_1-t_0}{N-1}, \ldots, \frac{t_1-t_0}{2}\right]$, where $t_0 \leq t \leq t_1$, the decoder $D_p$ takes all periodic parameters as its input, and maps back to the original motion curves:
\begin{equation}
\hat{\mathbf{L}}=f(\mathcal{T} ; \mathbf{A}, \mathbf{F}, \mathbf{B}, \mathbf{S})=\mathbf{A} \cdot \sin (2 \pi \cdot(\mathbf{F} \cdot \mathcal{T}-\mathrm{S}))+\mathbf{B}
\end{equation}
\begin{equation}
    \mathbf{\hat{G}_2}=h(\hat{\mathbf{L}})
\end{equation}

The network is learned with the periodic parameters via the following loss function, which forces the periodic parameters to reconstruct the original latent embedding $\mathbf{L}$.
\begin{equation}
\mathcal{L}_{phase}=\mathcal{L}_{phase-recon}(\mathbf{G}, \mathbf{\hat{G}_2})
\end{equation}

\begin{figure}
  \centering
  \begin{subfigure}{0.525\linewidth}
  \centering
    \includegraphics[width=0.6\textwidth,height=0.35\textwidth]{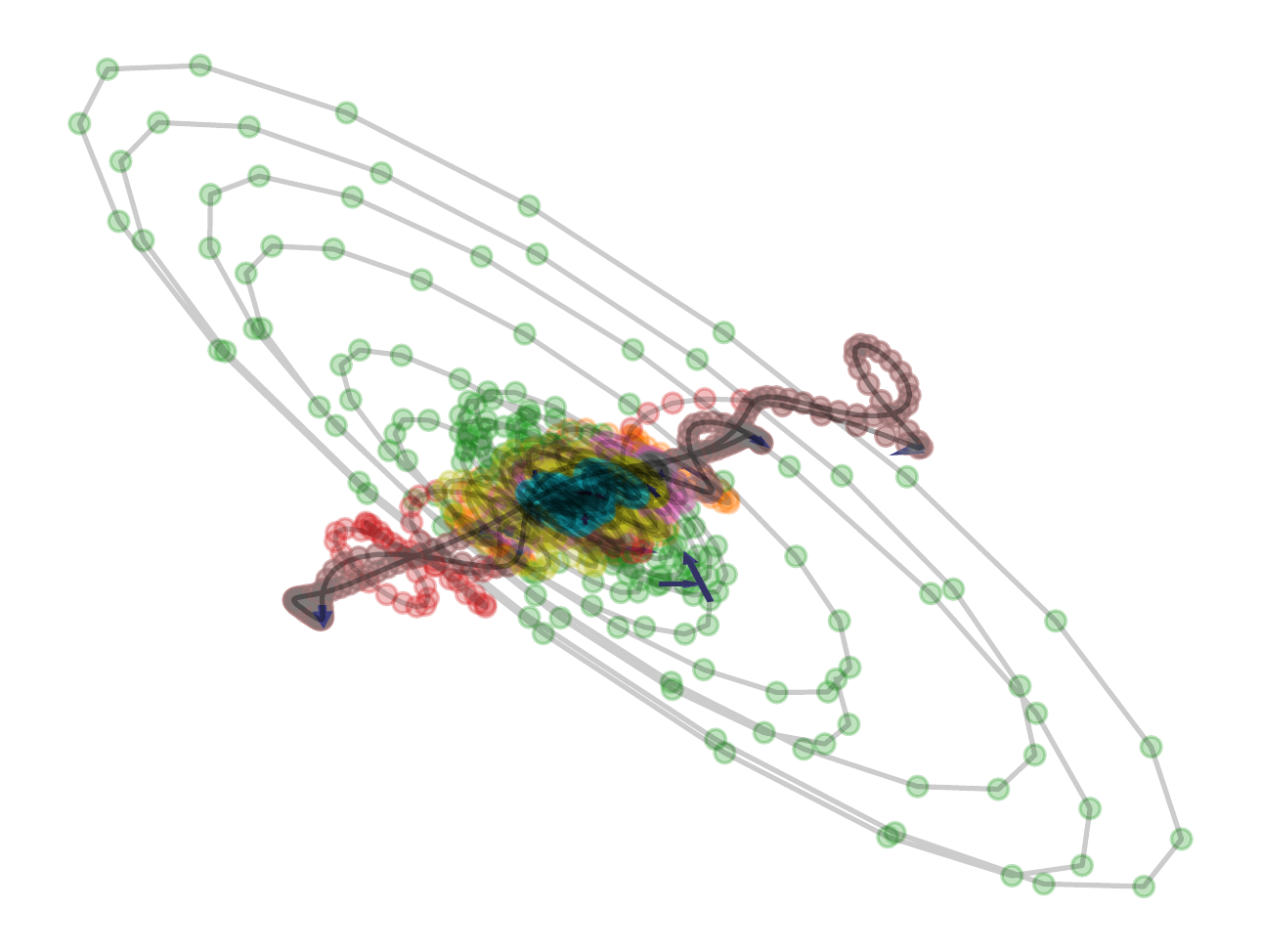}        
    \caption{Rotational velocity space.}
    \label{fig:short-a}
  \end{subfigure}
  \hfill
  \begin{subfigure}{0.9\linewidth}
  \centering
  \includegraphics[width=0.6\textwidth,height=0.35\textwidth]{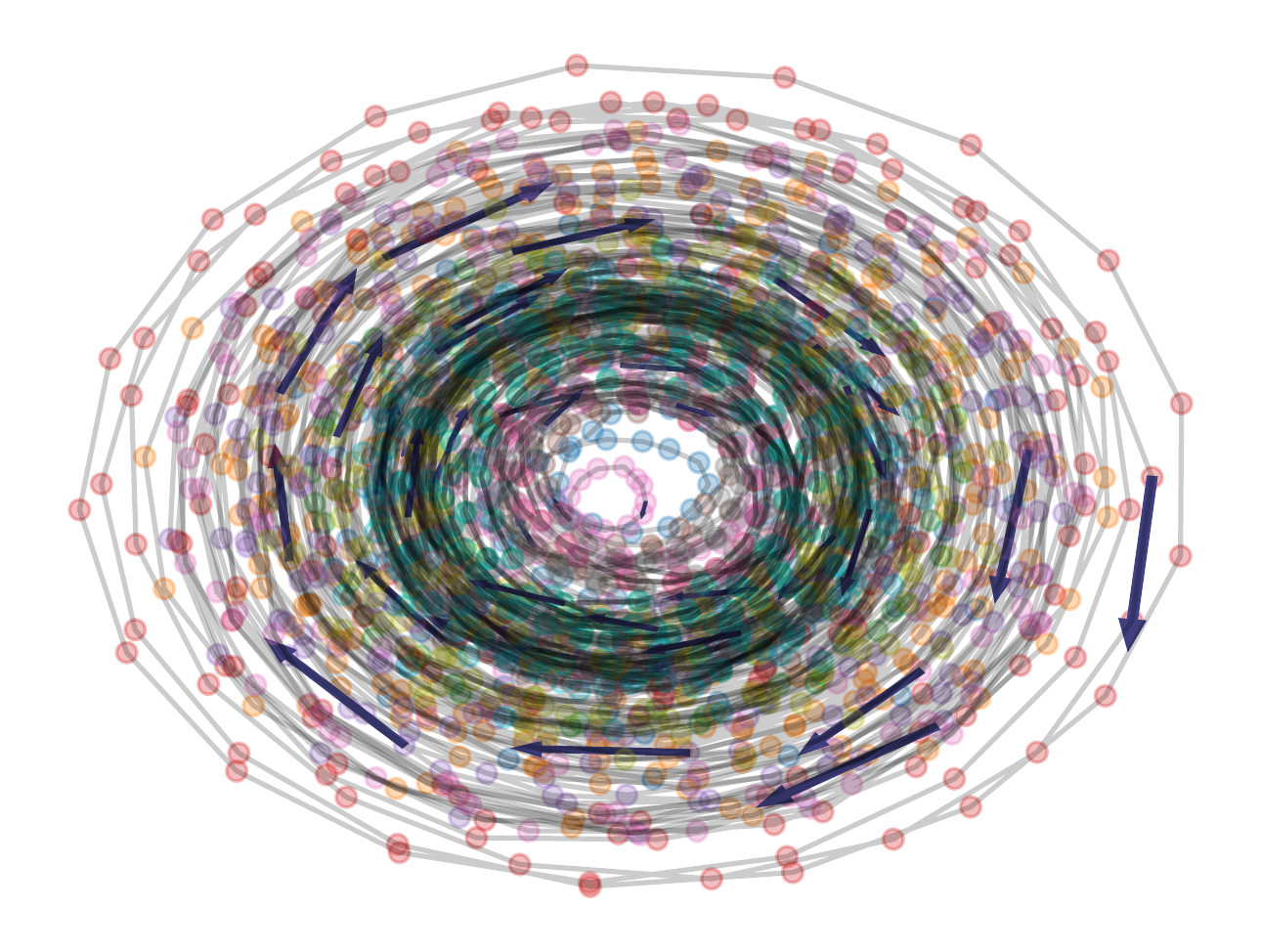}
    \caption{Latent space learned by convolutional and fully connected.}
    \label{fig:short-b}
  \end{subfigure}
    \hfill
  \begin{subfigure}{0.9\linewidth}
  \centering
  \includegraphics[width=0.6\textwidth,height=0.35\textwidth]{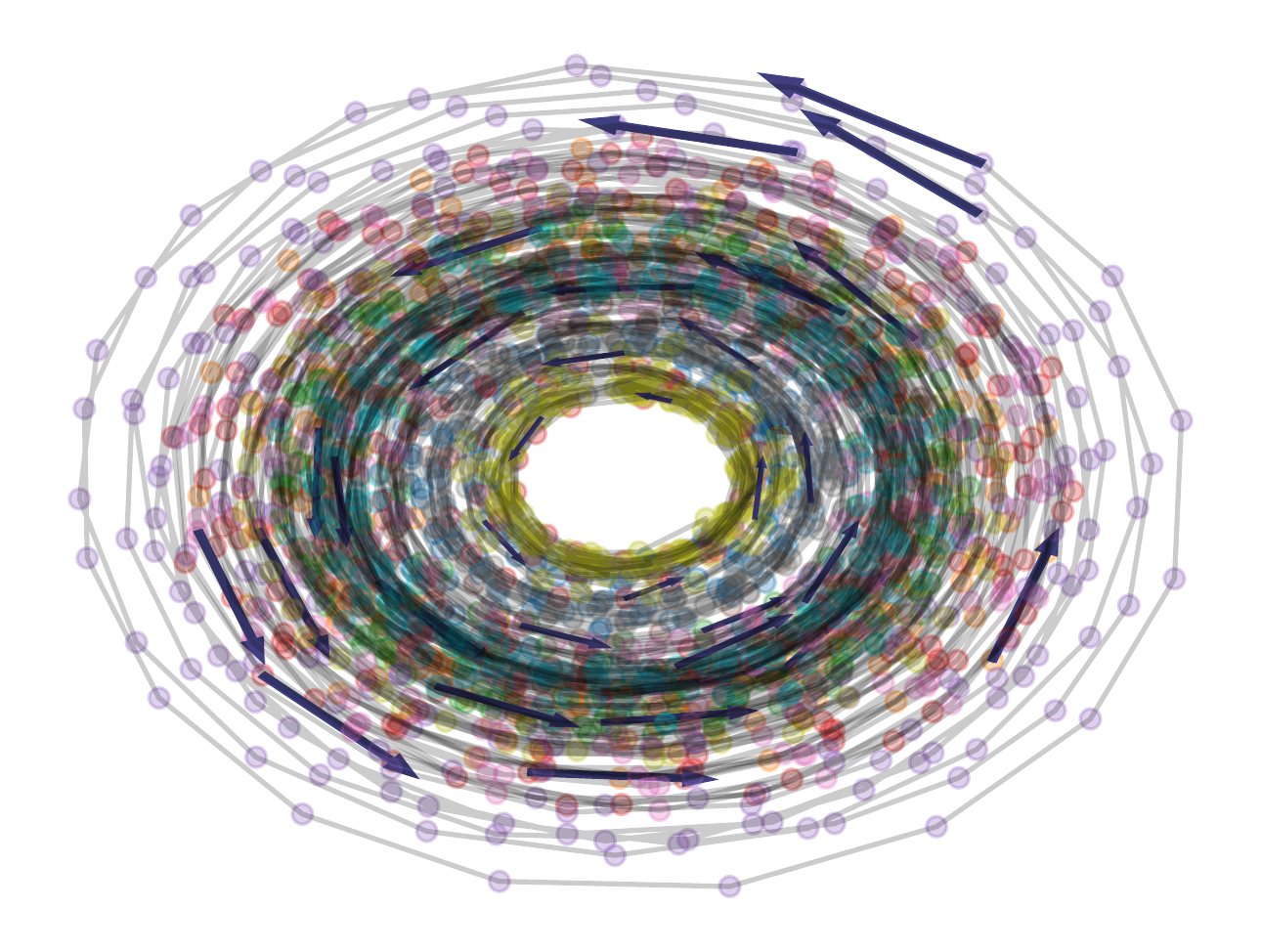}
    \caption{Phase space.}
    \label{fig:short-c}
  \end{subfigure}
  \caption{2D PCA embedding of feature distributions for different motion domains. Each color represents gestures from a single motion clip that are temporally connected, which means that neighboring frames in the motion data should be closely connected in the embedding}
  \label{fig:distributions}
\end{figure}

After training, phase manifold $\mathcal{P}$ of a sample motion which captures a lot of ``information" about the current state of the original time series data is
\begin{equation}
\mathcal{P}_{2 i-1}^{(t)}=\mathrm{A}_i^{(t)} \cdot \sin \left(2 \pi \cdot \mathrm{S}_i^{(t)}\right), \quad \mathcal{P}_{2 i}^{(t)}=\mathrm{A}_i^{(t)} \cdot \cos \left(2 \pi \cdot \mathrm{S}_i^{(t)}\right)
\end{equation}
where $\mathcal{P}\in \mathbb{R}^{2 M}$.

We visualize the phase features and the velocity features in Figure \ref{fig:distributions}.
The Principle Components (PCs) of the original joint rotational velocities, replacing the phase layers with fully connected layers and phase features are projected onto a 2D plane.
It can be observed that the phase manifold has a consistent structure similar to polar coordinates.
The cycles represent the primary period of the individual motions, where the timing is represented by the angle around the center, and the amplitude is the velocity of the motion. 
Samples smoothly transition between cycles of different amplitude or frequency, which indicate the transition between movements.

\begin{figure}[t]
  \centering
  \begin{subfigure}{0.49\linewidth}
  \centering
    \includegraphics[width=0.85\textwidth,height=1.0\textwidth]{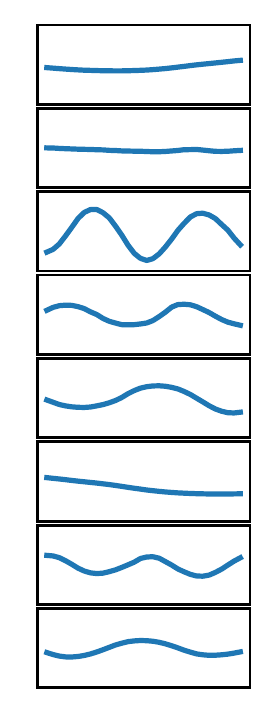}        
    \caption{The phase manifold of the seed code to be matched.}
    \label{fig:short-d}
  \end{subfigure}
  \hfill
  \begin{subfigure}{0.49\linewidth}
  \centering
  \includegraphics[width=0.85\textwidth,height=1.0\textwidth]{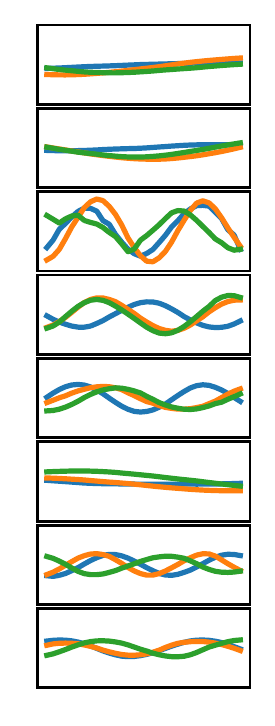}
    \caption{The phase manifold of the candidate to be matched.}
    \label{fig:short-e} 
  \end{subfigure}
  \caption{Sinusoidal diagram of learned phase manifold within a sliding window.
  The blue lines in (a) and (b) indicate the ground truth results, and the orange and green lines indicate the candidates to be matched, i.e., audio candidate $\mathbf{C}_a$ and text candidate $\mathbf{C}_t$.
  }
  \label{fig:phase manifold}
\end{figure}

The phase manifold of the motion curve is shown in Figure \ref{fig:phase manifold}.
Instead of using phase to set network weights as in \cite{DBLP:journals/tog/StarkeZKZ20, DBLP:journals/tog/ZhangSKS18, DBLP:journals/tog/HoldenKS17} to generate motion, we use phase guidance to select motion.
To find the appropriate motion alignments, we calculate the cosine similarity of the latter frames $N_{strid}$ of the manifold phase $\mathcal{P}_{-1}$ corresponding to the frame of the initial code and the first frames of the manifold phase $\mathcal{P}_{a/t}$ corresponding to the audio/text candidate $\mathbf{C}_a$ and $\mathbf{C}_t$ within $N_{phase}$ frames, which can be written as $concat[\mathcal{P}_{-1}^{[(N_{strid}-N_{phase}):]}, \mathcal{P}_{a/t}^{[N_{strid}:]}]$ and $concat[\mathcal{P}_{-1}^{[-N_{strid}:]}, \mathcal{P}_{a/t}^{[(N_{phase}-N_{strid}):]}]$.
Candidates with more similar phase manifolds will also have more natural motion as the final matching gestures.
Please refer to Appendix \ref{Supplementary-pseudo-code} for the pseudo-code of our algorithm.      

\section{Experiments}

\textbf{Dataset}.
We perform the training and evaluation on the BEAT dataset proposed in \cite{DBLP:journals/corr/abs-2203-05297}, which to our best knowledge is the largest publicly  available motion capture dataset for human gesture generation.
We divided the data into 8:1:1 by training, validation, and testing, and trained codebooks and baselines using data from all speakers.
Because motion matching is more time-consuming, we selected 4 hours of data from two speakers ``wayne'' (male speaker) and ``kieks'' (female speaker), and constructed separate databases for our experiments.

\textbf{Implementation Details.}
In this work, we use 15 joints corresponding to the upper body without hands or fingers.
3 × 3 rotation matrix features are computed as gesture sequences, with pose dimension $D_g$ is 9.
Down-sampling rate $d$ is 8.
The size $C_b$ of codebook $\mathcal{Z}_g$ is set to 512 with dimension $n_z$ is 512.
While training the gesture VQ-VAE, gesture data are cropped to a length of $T=240$ (4 seconds), using the ADAM \cite{DBLP:journals/corr/KingmaB14} optimizer (learning rate is e-4, $\beta_1$ = 0.5, $\beta_2$
= 0.98) with a batch size of 128 for 200 epochs.
We set $\beta$= 0.1 for Equation (\ref{beta}) and $\alpha_1$= 1, $\alpha_2$= 1 for Equation (\ref{alpha}).
In terms of motion matching, the window lengths for audio and text are 4 pose codes corresponding to each of the past and future speech information, with $d$=32.
As for phase guidance, we use the rotational velocity as input to the network.
We train the network using the AdamW \cite{DBLP:conf/iclr/LoshchilovH19} optimizer for 100 epochs, using a learning rate and weight decay both of $10^{-4}$ and a batch size of 128.
The number of phase channels $M$ is set to 8.
To calculate the phase similarity, the number of frames is set to $N_{phase}=8$ and $N_{stride}=3$.
The whole framework is learned in less than one day on one NVIDIA A100 GPU. 
The initial pose code is generated by randomly sampling a code from the codebook $\mathcal{Z}_g$.

\textbf{Evaluation Metrics.}
The distance between speed histograms is used to evaluate gesture quality.
We calculate speed-distribution histograms and compare to the speed distribution of natural motion by computing the Hellinger distance \cite{DBLP:conf/icmi/KucherenkoJWHAL20},
$H\left(\boldsymbol{h}^{(1)}, \boldsymbol{h}^{(2)}\right)=\sqrt{1-\sum_i \sqrt{h_i^{(1)} \cdot h_i^{(2)}}}$
between the histograms $\boldsymbol{h}^{(1)}$ and $\boldsymbol{h}^{(2)}$.
The Fréchet gesture distance (FGD) \cite{Yoon2020Speech} on feature space is proposed as a metric to quantify the quality of the generated gestures. This metric is based on the FID \cite{DBLP:conf/nips/HeuselRUNH17} metric used in image generation studies.
Similarly, we calculate FGD on raw data space used in \cite{DBLP:conf/emnlp/AhujaLIM20}.
To compute the FGD, we trained an autoencoder using the Trinity dataset \cite{IVA:2018}.
Lower Hellinger distance and FGD are better.
We also report average jerk, average acceleration, Canonical correlation analysis (CCA), Diversity, and Beat Align Score in Appendix \ref{Appendix-Objective}. 

\subsection{Comparison to Existing Methods}
\label{4.1}

\begin{table*}[t]
\centering
\caption{Quantitative results on test set.
Bold indicates the best metric.
Among compared methods, End2End \cite{DBLP:conf/icra/YoonKJLKL19}, Trimodal \cite{Yoon2020Speech}, StyleGestures \cite{DBLP:journals/cgf/AlexandersonHKB20}, KNN \cite{DBLP:conf/siggraph/HabibieESANNT22} and CaMN \cite{DBLP:journals/corr/abs-2203-05297} are reproduced using officially released code with some optimized settings.
For more details please refer to Appendix \ref{Appendix-Objective}.
Objective evaluation is recomputed using the officially updated evaluation code \cite{genea_numerical_evaluations}. Human-likeness and appropriateness are results of MOS with 95\% confidence intervals.
}
\label{tab:Quantitative}
\resizebox{0.88\textwidth}{21mm}{       
\begin{tabular}{cccccc}
\hline
\multirow{2}{*}{Name} & \multicolumn{3}{c}{Objective evaluation}                                                                                                                                                                   & \multicolumn{2}{c}{Subjective evaluation} \\ \cline{2-6} 
                      & \begin{tabular}[c]{@{}c@{}}Hellinger \\ distance average\end{tabular} $\downarrow$& \begin{tabular}[c]{@{}c@{}}FGD on \\ feature space\end{tabular} $\downarrow$& \begin{tabular}[c]{@{}c@{}}FGD on raw \\ data space\end{tabular} $\downarrow$& Human-likeness      & Appropriateness     \\ \hline
Ground Truth (GT)     & 0.0                                                                   & 0.0                                                             & 0.0                                                              &       3.79 $\pm$ 0.19              &       3.62 $\pm$ 0.21            \\
End2End \cite{DBLP:conf/icra/YoonKJLKL19}               & 0.146                                                                 & 64.990                                                          & 16739.978                                                        &   3.64 $\pm$ 0.11                 &   3.23 $\pm$ 0.14                 \\
Trimodal \cite{Yoon2020Speech}              & 0.155                                                                 & 48.322                                                          & 12869.98                                                         &    3.31 $\pm$ 0.17             &     3.20 $\pm$ 0.19               \\
StyleGestures \cite{DBLP:journals/cgf/AlexandersonHKB20}          & \textbf{0.136}                                                        & 35.842                                                          & 9846.927                                                         &  3.66 $\pm$ 0.08                  &  3.30 $\pm$ 0.11                 \\
KNN \cite{DBLP:conf/siggraph/HabibieESANNT22}                   & 0.364                                                                 & 43.030                                                          & 12470.061                                                        &   2.38 $\pm$ 0.10                  &  2.35 $\pm$ 0.13                   \\
CaMN \cite{DBLP:journals/corr/abs-2203-05297}       & 0.149 & 52.496    & 10549.455     & 3.65 $\pm$ 0.16       & 3.29 $\pm$ 0.15 \\
Ours    & \textbf{0.136}                                                        & \textbf{19.921}                                                 & \textbf{5742.281}                                                &   \textbf{4.00 $\pm$ 0.14 }                &   \textbf{3.66 $\pm$ 0.23}                  \\ \hline
\end{tabular}}
\end{table*}

We compare our proposed framework with End2End \cite{DBLP:conf/icra/YoonKJLKL19} (Text-based), Trimodal \cite{Yoon2020Speech} (Text, audio and identity, flow-based), StyleGestures \cite{DBLP:journals/cgf/AlexandersonHKB20} (Audio-based), KNN \cite{DBLP:conf/siggraph/HabibieESANNT22} (Audio, motion matching-based) and CaMN \cite{DBLP:journals/corr/abs-2203-05297} (Multimodal-based).
The quantitative results are shown in Table \ref{tab:Quantitative}.
According to the comparison, our proposed model consistently performs favorably against all the other existing methods on all evaluations.
Specifically, on the metric of Hellinger distance average, we achieve the same good results as StyleGestures.
Since well-trained models should produce motion with similar properties to a specific speaker, our method has a similar motion-speed profile for any given joint. 
And our method improves 15.921 ($44\%$) and 3837.068 ($39\%$) than the best compared baseline model StyleGestures on FGD on feature space and FGD on raw data space.

\textbf{User Study.}
To further understand the real visual performance of our method, we conduct a user study among the gesture sequences generated by each compared method and the ground truth data.
Following the evaluation in GENEA \cite{DBLP:journals/cgf/AlexandersonHKB20}, for each method, from the 30-minute test set we selected 38 short segments of test speech and corresponding test motion to be used in the evaluations. 
Segments are around 8 to 15 seconds long,  and ideally not shorter than 6 seconds.
The experiment is conducted with 23 participants separately. 
The generated gesture data is visualized on an avatar via Blender \cite{Blender} rendering.
For human-likeness evaluation, each evaluation page asked participants “How human-like does the gesture motion appear?” 
In terms of appropriateness evaluation, each evaluation page asked participants “How appropriate are the gestures for the speech?” 
Each page presented six videos to be rated on a scale from 5 to 1 with 1-point intervals with labels (from best to worst) “Excellent”, “Good”, “Fair”, “Poor”, and “Bad”. 
The mean opinion scores (MOS) on human-likeness and appropriateness are reported in the last two columns in Table \ref{tab:Quantitative}.

Our method significantly surpasses the compared state-of-the-art methods with both human-likeness and appropriateness, and even above the ground truth (GT) in human-likeness and appropriateness.
However, there is no significant difference compared to the appropriateness of GT.
According to the feedback from participants, our generated gesture is more “related to the semantics” with “more natural”, while our method “lacking power and exaggerated gestures” compared to GT.
More details regarding the user study are shown in Appendix \ref{appendix-User}.

\subsection{Ablation Studies}
\label{section:Ablation}
\begin{table*}[t]
\centering
\caption{Ablation studies results. `w/o' is short for `without'. Bold indicates the best metric.}
\label{tab:Ablation}
\resizebox{0.94\textwidth}{21mm}{
\begin{tabular}{cccccc}
\hline
\multirow{2}{*}{Name}                                                          & \multicolumn{3}{c}{Objective evaluation}                                                                                                                                                                   & \multicolumn{2}{c}{Subjective evaluation} \\ \cline{2-6} 
                                                                               & \begin{tabular}[c]{@{}c@{}}Hellinger \\ distance average\end{tabular} $\downarrow$ & \begin{tabular}[c]{@{}c@{}}FGD on \\ feature space\end{tabular} $\downarrow$& \begin{tabular}[c]{@{}c@{}}FGD on raw \\ data space\end{tabular} $\downarrow$& Human-likeness      & Appropriateness     \\ \hline
w/o wavvq + WavLM                                                              & 0.151                                                                 & 19.943                                                          & 6009.859                                                         &    3.87 $\pm$ 0.21                 &    3.64 $\pm$ 0.21                 \\
w/o audio                                                                      & 0.134                                                                 & 20.401                                                          & 5871.044                                                         &  3.87 $\pm$ 0.21                  & 3.63 $\pm$ 0.20                    \\
w/o text                                                                       & \textbf{0.118}                                                        & 23.929                                                          & 6389.866                                                         &    3.57 $\pm$ 0.29                 &  3.41 $\pm$ 0.23                   \\
w/o phase                                                                      & 0.138                                                                 & \textbf{19.195}                                                 & 5759.167                                                         &    3.90 $\pm$ 0.11                 &  3.65 $\pm$ 0.17                   \\
\begin{tabular}[c]{@{}c@{}}w/o motion matching\\ (GRU + codebook)\end{tabular} & 0.140                                                                 & 30.404                                                          & 11642.641                                                        &   3.78 $\pm$ 0.14                 & 3.43 $\pm$ 0.16                     \\
Ours                                                             & 0.136                                                                 & 19.921                                                          & \textbf{5742.281}                                                & \textbf{4.07 $\pm$ 0.15}           &   \textbf{ 3.77 $\pm$ 0.21 }              \\ \hline
\end{tabular}}
\end{table*}

Moreover, we conduct ablation studies to address the performance effects of different components in the framework.
The results of our ablations studies are summarized in Table \ref{tab:Ablation}.
The visual comparisons of this study can be also referred to the supplementary video.

We explore the effectiveness of the following components: 
(1) Levenshtein distance,
(2) audio modality,
(3) text modality,
(4) phase guidance,
(5) motion matching.
We performed the experiments on each of the five components, respectively.

Supported by the results in Table \ref{tab:Ablation}, when we do not use vq-wav2vec or Levenshtein distance to measure the similarity of corresponding speech of gestures, but use WavLM \cite{DBLP:journals/jstsp/ChenWCWLCLKYXWZ22} pre-trained on Librispeech and cosine similarity instead, the performances of all metrics have deteriorated. 
The change of FGD on feature space was not significant.
The Hellinger distance average and FGD on raw data space deteriorated by 0.015 (11\%), and 267.578 (4.7\%), respectively.
When the model is trained without audio, we select two candidates for text-based motion matching instead of one and then synthesize the gesture based on phase guidance.
The FGD on feature space and FGD on raw data space deteriorated by 0.48 (2\%) and 128.763 (2\%), respectively.
When the model is trained without text, similarly, two candidates for audio-based motion matching are selected instead of one.
The FGD on feature space and FGD on raw data space deteriorated by 4.008 (20\%) and 647.585 (11\%), respectively.
Notice that when one of the modalities of both text and speech is not used, the FGD metric increases while the Hellinger distance average metric decreases, which indicates that the quality of the gestures generated when only one modality is used decreases, but the distribution of the velocity becomes better.
\begin{figure}[t]
  \centering
   \includegraphics[width=\linewidth]{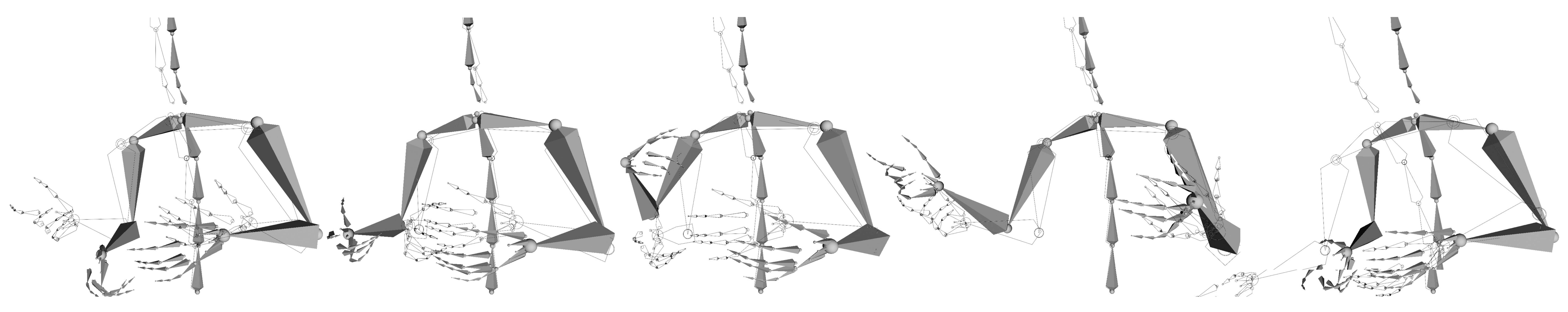}
   \caption{Comparison of gestures blended with and without phase guidance. Black shading indicates no phase guidance.}
   \label{fig:Comparison}
\end{figure}
When the phase guidance is removed, we select one candidate every time between two candidates using the distance in the rotational velocity space (Not randomly select one candidate).
The results showed a slight increase in Hellinger distance average and FGD on raw data space and a slight decrease in FGD on feature space, but none of the changes were significant.
One possible explanation is that the phase space loses information from the original feature space to generate more natural actions, as shown in Figure \ref{fig:Comparison}.
When the model is trained using deep gated recurrent unit (GRU) \cite{DBLP:conf/emnlp/ChoMGBBSB14} to learn pose code instead of motion matching,
The FGD on feature space and FGD on raw data space deteriorated by 10.483 (53\%) and 6460.202 (107\%), respectively.
This demonstrates the advantage of the matching model over the generative model.
Further, this model has a comparable performance with baselines, which also proves the efficiency of the codebook encoded gesture space.

\textbf{User Study.}
Similarly, we conduct a user study of ablation studies.
We use the same approach as in Section \ref{4.1}, with the difference that we use another avatar character  to test the robustness of the results.
The MOS on human-likeness and appropriateness are shown in the last two columns in Table \ref{tab:Ablation}.
The score of our proposed framework is similar to the previous one, which will be a bit higher, indicating that even if the generated results are the same, the rating may be related to the visual perception of different character.
However, there was no significant change.
The results demonstrate that the final performance of the model decrease without any module. 
Notice that the score without text decreases more than the score without audio, indicating that the matched gestures are mostly semantically related.
Not using phase space has the least effect on the results, which is consistent with expectations, since phase only provides guidance.
Another significant result is that the results without audio and using audio without Levenshtein distance or audio quantization are close, which effectively indicates the effectiveness of Levenshtein distance.
Furthermore, the results without model matching has a comparable performance in terms of naturalness with GT, the effectiveness of the codebook encoded gesture unit was also confirmed.

\subsection{Controllability}

\begin{figure}[t]
  \centering
   \includegraphics[width=0.6375\linewidth]{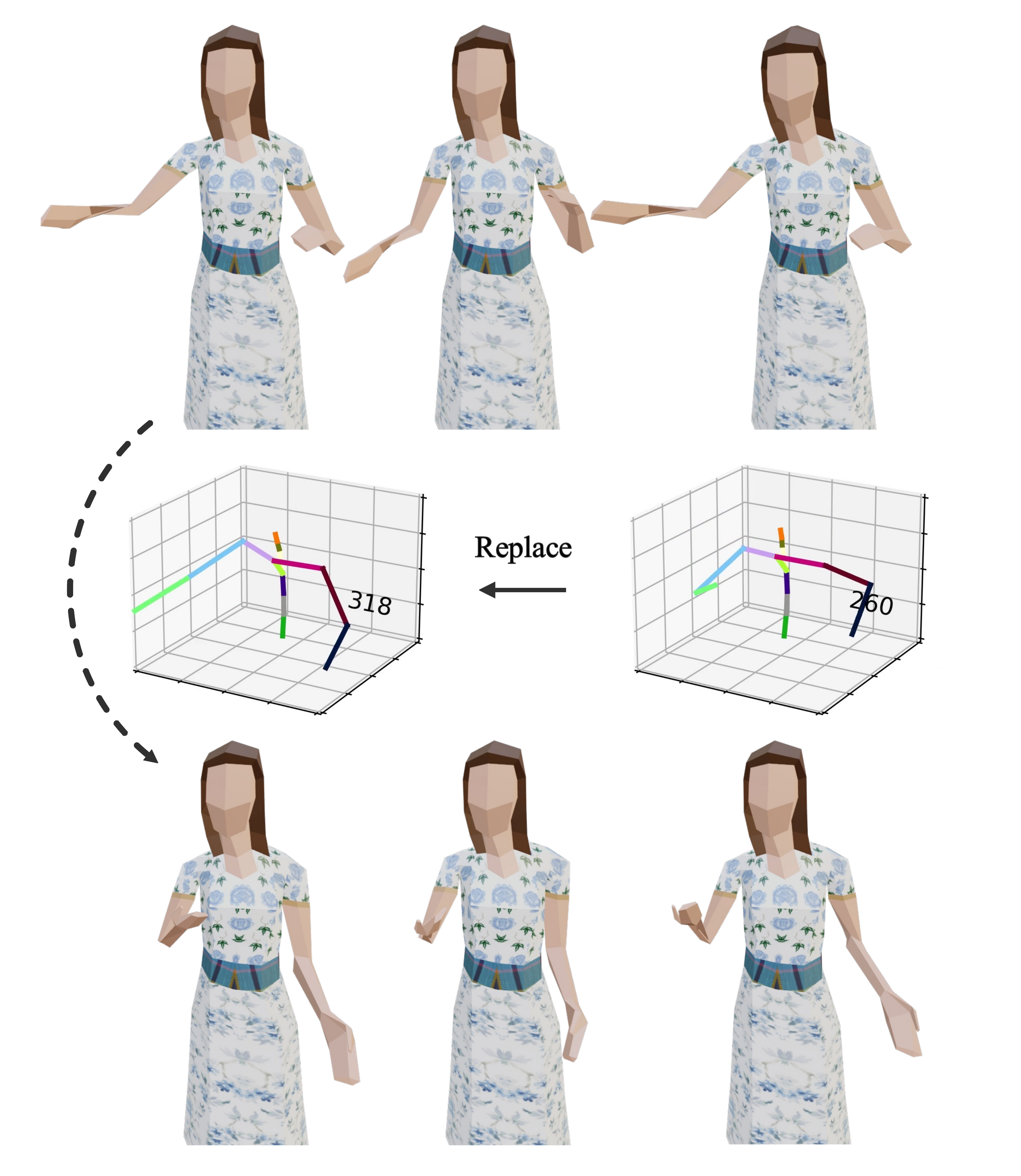}
   \caption{Visualization of how a gesture changes when codes are changed. We replace twelve code `318' in a motion sequence (240 frames, 30 codes) with code `260' on a habitual right-handed speaker.}
   \label{fig:Controllability}
\end{figure}

Since we use motion matching to generate gestures, it is easy to control the gesture or take out the code for interpretation.
For example, to generate a sequence of body gestures where the left wrist is always above a specified threshold \textbf{r}, the search can be restricted to consider only codes corresponding to wrists above \textbf{r}. 
Here is an example. 
The most frequent gesture in the database of ``wayne" that we found, besides the average pose, is the gesture corresponding to the code `318', as shown in Figure \ref{fig:Controllability}. 
This can be explained by that ``wayne" is a habitual right-handed speaker.
We choose a preferred left-handed code `260' to replace it, and the result is shown in the figure.
In practice, to produce natural gestures, it is sufficient to add the code frequency score and adjust the weights appropriately when matching.
Please refer to the supplementary video for more comparisons.

\section{Discussion and Conclusion}

In this paper, we present a quantization-based and phase-guided motion matching framework for speech-driven gesture generation. Specifically, we  address the random jittering issue by using discrete representation that encodes human gestures.
Besides, we tackle the inherent asynchronicity of speech and gestures and flexibility of the current motion matching models by Levenshtein distance based on audio quantization.
Then, phase-guided audio-based or text-based candidates are used as the final result.
Experiments on the standard benchmark (i.e., BEAT dataset) along with user studies show that proposed framework achieves state-of-the-art performance both qualitatively and quantitatively.
There is room for improvement in this research, besides text and audio, more modalities (e.g. emotions, facial expressions) could be taken into consideration to generate more appropriate gestures.

\section*{Acknowledgments}

This work is supported by National Natural Science Foundation of China (NSFC) (62076144), Shenzhen Key Laboratory of next generation interactive media innovative technology (ZDSYS20210623092001004) and Shenzhen Science and Technology Program (WDZC20220816140515001, JCYJ20220818101014030).



{\small
\bibliographystyle{ieee_fullname}
\bibliography{egbib}
}

\clearpage

\renewcommand\thesection{\Alph{section}}

\setcounter{section}{0}
\section{The design choice of our method}
A gesture motion is composed of a sequence of gesture units, such as swiping hands from left to right and holding hands at a position \cite{DBLP:journals/corr/abs-2210-01448}. We design our algorithm according to this observation, paying special attention to the construction and selection of gesture units. Specifically, due to the good performance of VQ-VAE in quantization, we trained a gesture VQ-VAE for 200 epochs to mine these gesture units from the dataset, similar to existing works \cite{DBLP:journals/corr/abs-2210-01448, DBLP:journals/corr/abs-2207-01696, DBLP:journals/tog/HongZPCYL22, lucas2022posegpt} \citeA{DBLP:journals/speech/SadoughiB19}. In our settings, each code corresponds to one gesture unit 
which is 8 ($d$) frames of gesture motions. 
Unlike Bailando \cite{siyao2022bailando}, Gensture2Vec \citeA{9981117} and VQ-Text2Sign \citeA{xie2022vector} using position as input features, our VQ-VAE is trained with rotation instead, which can represent the motion better.
 To find gesture candidates that match a given piece of audio and corresponding text, 
we quantize the audio first, 
because our ablation studies in Table \ref{tab:Ablation} illustrate that the Levenshtein distance based on discrete audio alleviates the inherent asynchrony problem of gesture and audio, and achieves better results than the non-discrete counterpart.
We don't need to quantize the text, since it is discrete already.

In terms of audio quantization, the audio is represented by two groups containing 320 tokens, for a total of $320^2$ results, or 102.4K tokens. 
\textbf{G} refers to the motion sequence (position, velocity, acceleration, rotation, Euler angles, quaternions, etc.). We use rotation for VQ-VAE, and rotation velocity for periodic autoencoder.
For motion matching, we first calculate $\hat{\mathbf{C}}_{a}$ and $\hat{\mathbf{C}}_{t}$ based on the audio and text.
We also calculate the distance between all gesture codes in codebook and the previous pose code $\mathbf{g}_{-1}$ to obtain $\hat{\mathbf{C}}_{g}$ for motion coherence. 
Then $\hat{\mathbf{C}}_{g}$ and $\hat{\mathbf{C}}_{a}$ determine audio-based candidate and $\hat{\mathbf{C}}_{g}$ and $\hat{\mathbf{C}}_{t}$ determine text-based candidate.
 The final gesture is selected according to the continuity of the phase of the previous gesture and the phase of the two candidate gestures.
 In practice, we find that if all the codes of the codebook are action-matched, speech-matched, and text matched with the same probability, sometimes there will be some strange gestures because some codes appear very infrequently and the database set is so large that sometimes the gesture corresponding to this speech has not yet started/ended. So in practice in motion matching we add a small weight (0.05) to the probability of codebook occurrence ranking for constraint, to avoid gestures that are almost not in the database to exist in the inference process. Please see the code for detailed implementation.

\section{Proposed Algorithm}
\label{Supplementary-pseudo-code}

\begin{algorithm}[!htp]
  \caption{QPGesture search}
  \KwData{database contains quantized audio, quantized gesture, context, and phase}
  \KwIn{a discrete text sequence $\mathbf{t}=[\mathbf{t}_0, \mathbf{t}_1, \dots, \mathbf{t}_{T^{\prime}-1}]$, 
  a discrete audio sequence $\mathbf{a}_{\mathbf{q}}=[\mathbf{a}_{\mathbf{q}, 0}, \mathbf{a}_{\mathbf{q}, 1}, \dots, \mathbf{a}_{\mathbf{q}, T^{\prime}-1}]$, initial pose code $\mathbf{g}_{-1}$, initial phase $\mathcal{P}_{-1}$, $k \in \mathbb{Z}$, the desired k-best candidates, control masks $\mathbf{M} = [\mathbf{m}_0, \mathbf{m}_1, \dots, \mathbf{m}_{T^{\prime}-1}]$ (optional)}
  \KwOut{$\hat{\mathbf{G}_o}= [\hat{\mathbf{G}}_{o,0}, \hat{\mathbf{G}}_{o,1} \dots, \hat{\mathbf{G}}_{o,T^{\prime}-1}]$}
  t = 0, codebook size $C_b$
  
  initialize $\hat{\mathbf{G}_o} = [\mathbf{g}_{-1}]$, $\hat{\mathcal{P}_o} = [\mathcal{P}_{-1}]$, 
  
  \While{$t$ {\bf in} len(testing dataset)}
  {
    c\_dist = []$\times C_b$, c\_a = []$\times C_b$, c\_t = []$\times C_b$
    
    a\_dist = [$\infty$]$\times C_b$, t\_dist = [$\infty$]$\times C_b$

    \For{$code=0; code < C_b$}
    {
        $c\_dist[code] = d(D_g\left(\hat{\mathbf{G}_o}[-1]\right), D_g\left(code\right))$
    }
    
    \For{$s=0;s < \text{len}(database)$}
    {
    \For{code {\bf in} database$[s]$}
    {
        \If{$\mathbf{m}_s$ {\rm is not masked}}
        {
            \If{$d({quantized\ audio}[s][code])<a\_dist[code]$}
            {
            $a\_dist[code]=d({quantized\ audio}[s][code])$
            
            $c\_a[code] = quantized\ audio[s][code:code+step size -1]$
            }
            \If{$d({context}[s][code])<t\_dist[code]$}
            {
            $t\_dist[code]=d({context}[s][code])$
            
            $c\_t[code] = context[s][code:code+step size -1]$
            }
        }
    }
    }
    
    $R_{c} = relrank(c\_dist)$,
    $R_{a} = relrank(a\_dist)$,
    $R_{t} = relrank(t\_dist)$
    
    $R_{c,a}=R_{c}+R_{a}$ (elem. wise)
    
    $R_{c,t}=R_{c}+R_{t}$ (elem. wise)
    
    sort $R_{c,a}$, sort its indices into $I_{c,a}$
    
    sort $R_{c,t}$, sort its indices into $I_{c,t}$
    
    $\hat{\mathbf{C}}_{a,t}=I_{c,a}[k]$, $\hat{\mathbf{C}}_{t,t}=I_{c,t}[k]$
    
    \If{$d(concat[\hat{\mathcal{P}_{o}}[-1]^{[(N_{strid}-N_{phase}):]}, \mathcal{P}_{a,t}^{[N_{strid}:]}]$, $concat[\hat{\mathcal{P}_{o}}[-1]^{[-N_{strid}:]}, \mathcal{P}_{a,t}^{[(N_{phase}-N_{strid}):]}]) <$ 
    $d(concat[\hat{\mathcal{P}_{o}}[-1]^{[(N_{strid}-N_{phase}):]}, \mathcal{P}_{t,t}^{[N_{strid}:]}]$, $concat[\hat{\mathcal{P}_{o}}[-1]^{[-N_{strid}:]}, \mathcal{P}_{t,t}^{[(N_{phase}-N_{strid}):]}])$}
    {append($\hat{\mathbf{G}_o}, \hat{\mathbf{C}}_{a,t}$),
    append($\hat{\mathcal{P}_o}, \hat{\mathcal{P}}_{a,t}$)
    }
    \Else
    {
    append($\hat{\mathbf{G}_o}, \hat{\mathbf{C}}_{t,t})$,
    append($\hat{\mathcal{P}_o}, \hat{\mathcal{P}}_{t,t})$
    }
  }
  return $\hat{\mathbf{G}_o}[1:]$
\end{algorithm}

A more detailed and procedural description of our proposed QPGesture approach is shown in Algorithm 1.

\section{Dataset and processing.}
 We chose BEAT dataset because to our knowledge it is the largest publicly available motion capture dataset.
And we will add more results of the baseline model for comparison later.

Since 2D datasets converted to 3D coordinates (pseudo GT) are low quality that are difficult to use, we plan to add more experiments on other motion capture datasets.
 Even based on motion capture, the hand quality of most datasets is still low \citeA{nyatsanga2023comprehensive}.
Datasets claimed with high-quality hand motion capture were still reported to have poor hand motion, 
e.g., ZEGGS Dataset in \citeA{DBLP:journals/corr/abs-2211-09707} and Talking With Hands in \cite{DBLP:conf/icmi/YoonWKVNTH22}. 
We found the hand quality of BEAT is not good enough, especially when retargeted to an avatar, so we ignore hand motion currently, and leave it to future work.

\section{Details of Baseline Implementation}

We used the 15 joints of the upper body(spine, spine1, spine2, spine3, head, neck, neck1, L/R shoulders, L/R
arms, and L/R forearms, L/R hands).
The gestures for all models were at 60 frames per second (fps). 
Because we found that using a pre-trained model to extract features was better than using 1D convolution,
for Trimodal \cite{Yoon2020Speech}, we used WavLM features instead of the original 1D convolution, while aligning the temporal dimensions using linear interpolation. 
For KNN \cite{DBLP:conf/siggraph/HabibieESANNT22}, we found that changing the step size from 2 frames at the original 15 fps to 30 frames at 60 fps had comparable results. 
However, we found that generating fake gestures for training the GAN in the second stage without overlapping frames and with 5 frames as the step size takes several months, which is intolerable. 
This could be due to 1) a large amount of data in the BEAT dataset itself,
2) the significant increase in the number of frames at 60 fps, and 3) the time-consuming KNN search itself (the time complexity of KNN search is O($n^4$) compared to time complexity of O($n^2$) of our method using audio quantization and gesture quantization). So we used mismatched gestures instead of KNN-matched gestures with 50\% likelihood from top2-top15 in the original KNN method as the gestures used for training the GAN in the second stage.
For CaMN \cite{DBLP:journals/corr/abs-2203-05297}, at the time we used the BEAT dataset, facial modality was not yet available\footnote{https://pantomatrix.github.io/BEAT-Dataset/}, so we used text, speech, speaker identity, and emotion as inputs to the CaMN network.

\section{Objective evaluation}
\label{Appendix-Objective}
\subsection{Evaluation Metrics}
\textbf{Average jerk and Acceleration.} The third and second time derivatives of the joint positions are called jerk and acceleration \citeA{10.1145/3308532.3329472}, respectively.
The average of these two metrics is usually used to evaluate the smoothness of the motion.
A natural system should have the average jerk and acceleration similar to natural motion.

\textbf{Canonical Correlation Analysis.}
The purpose of Canonical correlation analysis (CCA) \citeA{DBLP:journals/speech/SadoughiB19} is to project two sets of vectors into a joint subspace and then find a sequence of linear transformations of each set of variables that maximizes the relationship between the transformed variables. 
CCA values can be used to measure the similarity between the generated gestures and the real ones.
The closer the CCA to 1, the better.

\textbf{Diversity and Beat Align Score.}
We use the method in \cite{DBLP:conf/iccv/0071KPZZ0B21} to calculate the beats of audio, and follow \cite{siyao2022bailando} to calculate the beats and diversity of gesture.
The greater these metrics are, the better.

\subsection{Objective Evaluation Results}

\begin{table*}[!t]
\centering
\caption{
Quantitative results on test set.
Bold indicates the best metric, i.e. the one closest to the ground truth.}
\label{add-obj-baselines}
\scalebox{0.95}{
\begin{tabular}{cccccccc}
\hline
Name &
  \begin{tabular}[c]{@{}c@{}}Average \\ jerk\end{tabular} &
  \begin{tabular}[c]{@{}c@{}}Average \\ acceleration\end{tabular} &
  \begin{tabular}[c]{@{}c@{}}Global \\ CCA\end{tabular} &
  \begin{tabular}[c]{@{}c@{}}CCA for \\ each sequence\end{tabular} &
  \begin{tabular}[c]{@{}c@{}}Diversity on \\ feature space \end{tabular} $\uparrow$&
  \begin{tabular}[c]{@{}c@{}}Diversity on \\ raw data space \end{tabular} $\uparrow$ &
  \begin{tabular}[c]{@{}c@{}}Beat Align \\ Score \end{tabular} $\uparrow$ \\ \hline
Ground Truth  & 996.32 $\pm$ 235.86       & 31.89 $\pm$ 6.80          & 1.000          & 1.00 $\pm$ 0.00  & 2.81          & 50.87        & 0.2064          \\
End2End \cite{DBLP:conf/icra/YoonKJLKL19}       & 143.68 $\pm$ 10.45        & 7.09 $\pm$ 0.34           & 0.429          & \underline{0.72 $\pm$ 0.14}  & 1.45          & 20.82          & \underline{0.2370}        \\
Trimodal \cite{Yoon2020Speech}     & 157.87 $\pm$ 12.08        & 7.98 $\pm$ 0.53           & 0.807          & \textbf{0.74 $\pm$ 0.19} & 1.91          & 17.21          & 0.1221          \\
StyleGestures \cite{DBLP:journals/cgf/AlexandersonHKB20} & \underline{280.44 $\pm$ 21.43}           & \textbf{23.58 $\pm$ 7.21} & 0.953          & 0.71 $\pm$ 0.12 & \textbf{5.80} & \textbf{29.88} & 0.1871          \\
KNN \cite{DBLP:conf/siggraph/HabibieESANNT22}           & \textbf{423.83 $\pm$ 100.10} & \underline{40.77 $\pm$ 8.12}          & \textbf{0.998} & 0.63 $\pm$ 0.21 & 3.23          & 19.42          & 0.2009          \\
CaMN \cite{DBLP:journals/corr/abs-2203-05297}          & 159.54 $\pm$ 13.99 & 8.96 $\pm$ 0.55 & 0.626 & 0.70 $\pm$ 0.17 &  2.26             &  18.60              &  0.1489\\
Ours          & 182.11 $\pm$ 18.15           & 9.87 $\pm$ 0.66           & \underline{0.985}          & 0.69 $\pm$ 0.14 & \underline{4.05}          & \underline{23.13}          & \textbf{0.2557} \\ \hline
\end{tabular}
}
\end{table*}

\begin{table*}[!t]
\centering
\caption{Ablation studies results. 
`w/o' is short for `without'.
Bold indicates the best metric, i.e. the one closest to the ground truth.}
\label{add-obj-Ablation}
\scalebox{0.9}{
\begin{tabular}{cccccccc}
\hline
Name &
  \begin{tabular}[c]{@{}c@{}}Average \\ jerk\end{tabular} &
  \begin{tabular}[c]{@{}c@{}}Average \\ acceleration\end{tabular} &
  \begin{tabular}[c]{@{}c@{}}Global \\ CCA\end{tabular} &
  \begin{tabular}[c]{@{}c@{}}CCA for \\ each sequence\end{tabular} &
  \begin{tabular}[c]{@{}c@{}}Diversity on \\ feature space \end{tabular} $\uparrow$&
  \begin{tabular}[c]{@{}c@{}}Diversity on \\ raw data space \end{tabular} $\uparrow$ &
  \begin{tabular}[c]{@{}c@{}}Beat Align \\ Score \end{tabular} $\uparrow$ \\ \hline
Ground Truth (GT)     & 996.32 $\pm$ 235.86       & 31.89 $\pm$ 6.80          & 1.000          & 1.00 $\pm$ 0.00  & 2.81          & 50.87        & 0.2064      \\                               
w/o wavvq + WavLM                                                              & 168.09 $\pm$ 22.44                  & 9.18 $\pm$ 0.81                             & \textbf{0.993}                 & 0.69 $\pm$ 0.13  & \underline{8.49} & 18.82 & 0.2098                            \\
w/o audio                                                                      & 176.84 $\pm$ 14.61                  & 9.60 $\pm$ 0.50                             & \textbf{0.993}                 & 0.68 $\pm$ 0.13 & 8.42 & \textbf{25.83} &  0.2001                            \\
w/o text                                                                       & \textbf{196.61 $\pm$ 29.34}         & \textbf{10.68 $\pm$ 1.22}                   & 0.961                          & 0.71 $\pm$ 0.15 & 7.53 & 15.78 & 0.1699                             \\
w/o phase                                                                      & 176.94 $\pm$ 21.41                  & 9.60 $\pm$ 0.80                             & 0.986                          & \underline{0.72 $\pm$ 0.13}   & 4.83 & 15.30 & \textbf{0.3076}                           \\
\begin{tabular}[c]{@{}c@{}}w/o motion matching\\ (GRU + codebook)\end{tabular} & 141.52 $\pm$ 9.65                   & 7.56 $\pm$ 0.56                             & 0.694                          & \textbf{0.75 $\pm$ 0.14}   & \textbf{10.98} & 12.51 & 0.2303                \\
Ours                                                                           & \underline{182.11 $\pm$ 18.15}                  & \underline{9.87 $\pm$ 0.66}                            & 0.985                          & 0.69 $\pm$ 0.14  & 4.05          & \underline{23.13}          & \underline{0.2557}                            \\ \hline
\end{tabular}
}
\end{table*}

We used Trinity dataset to calculate FGD because both Trinity and BEAT are captured with Vicon, having the same names and number of joints, as in \cite{Yoon2020Speech}.
The results of our additional objective evaluation compared to the existing model are shown in Table \ref{add-obj-baselines}.
From the results, we can observe that KNN performs better than our proposed framework on three metrics: average jerk, average acceleration and global CCA. 
StyleGestures performs best on Average acceleration. And Trimodal has the best performance on CCA for each sequence.
We can see that our model is the best match to the beats of the audio, but not as good as StyleGesture in terms of diversity.
The video results show that StyleGesture has a lot of cluttered movements, increasing diversity while decreasing human-likeness and appropriateness.

The results of additional objective evaluations of our ablation studies are shown in Table \ref{add-obj-Ablation}.
When we do not use vq-wav2vec or Levenshtein distance to measure the similarity of corresponding speech of gestures, but use WavLM and cosine similarity instead, the average jerk and average acceleration are worst.
When the framework is inferenced without text, the average jerk, average acceleration and CCA for each sequence are better, but the global CCA is decreased.
When the model is trained using deep gated recurrent unit (GRU) to learn pose code instead of motion matching, the best CCA for each sequence is obtained.
For diversity, more diverse may indicate a more clutter-free gesture; and for scores, a better match with rhythm does not indicate a better semantic match. These objective measures are not consistent with subjective scoring.

However, this is consistent with current human subjective perception \cite{DBLP:conf/iui/KucherenkoJYWH21, DBLP:conf/icmi/YoonWKVNTH22} that speech-driven gestures lack proper objective metrics, even for FGD \citeA{DBLP:journals/corr/abs-2212-04495}.
Current research on speech-driven gestures prefers to conduct only subjective evaluation \citeA{DBLP:journals/corr/abs-2211-09707}.
In conclusion, we would like to emphasize that objective evaluation is currently not particularly relevant for assessing gesture generation \cite{DBLP:conf/iui/KucherenkoJYWH21}. 
Subjective evaluation remains the gold standard for comparing gesture generation models \cite{DBLP:conf/iui/KucherenkoJYWH21}.

\section{User Study}
\label{appendix-User}
Segments should be more or less complete phrases, starting at the start of a word and ending at the end of a word.
We made sure there were no spoken phrases that ended on a ``cliffhanger'' in the evaluation.
The user study was conducted by subjects with good English proficiency.
The reward is about 7.5 USD each person, which is about the average wage level \cite{DBLP:conf/icmi/YoonWKVNTH22}.
More detailed demographic data of the subjects who participated in the subjective evaluation are as follows.
\begin{itemize}
    \item Gender: Participants were approximately 90\% were male and 10\% were female.
    \item Region: They were overwhelmingly residents of mainland China, and one international student from Malaysia. They are all students from our lab\footnote{https://thuhcsi.github.io/labintro.html}.
    \item Age: All participants were between the ages of 20-28.
\end{itemize}
The questions for user study follow GENEA 2022 \cite{DBLP:conf/icmi/YoonWKVNTH22}.
If there is no overlap in the 95\% confidence intervals of the ratings between the different models, then the difference is considered to be statistically significant.

\begin{figure}[t]
  \centering
  \includegraphics[width=0.39\textwidth,height=0.57\textwidth]{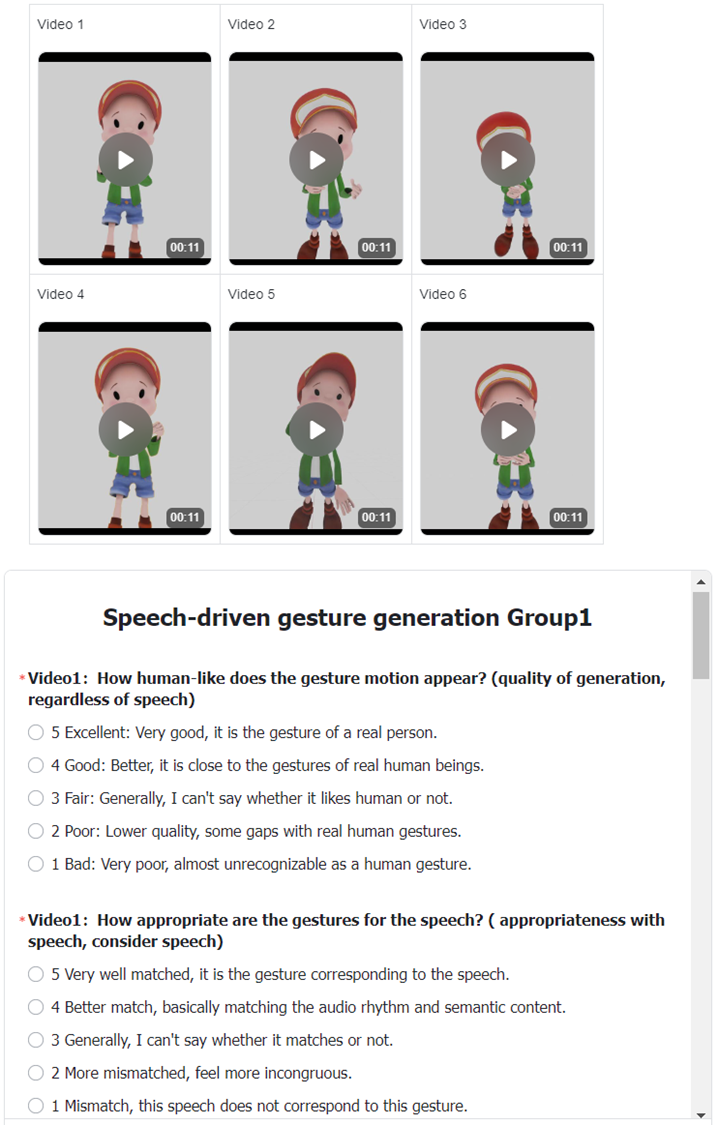}
  \caption{Screenshot of the parallel rating interface from the user study for comparison with existing methods.}
  \label{fig:comparison with existing methods}
\end{figure}

\begin{figure}[t]
  \centering
    \includegraphics[width=0.39\textwidth,height=0.57\textwidth]{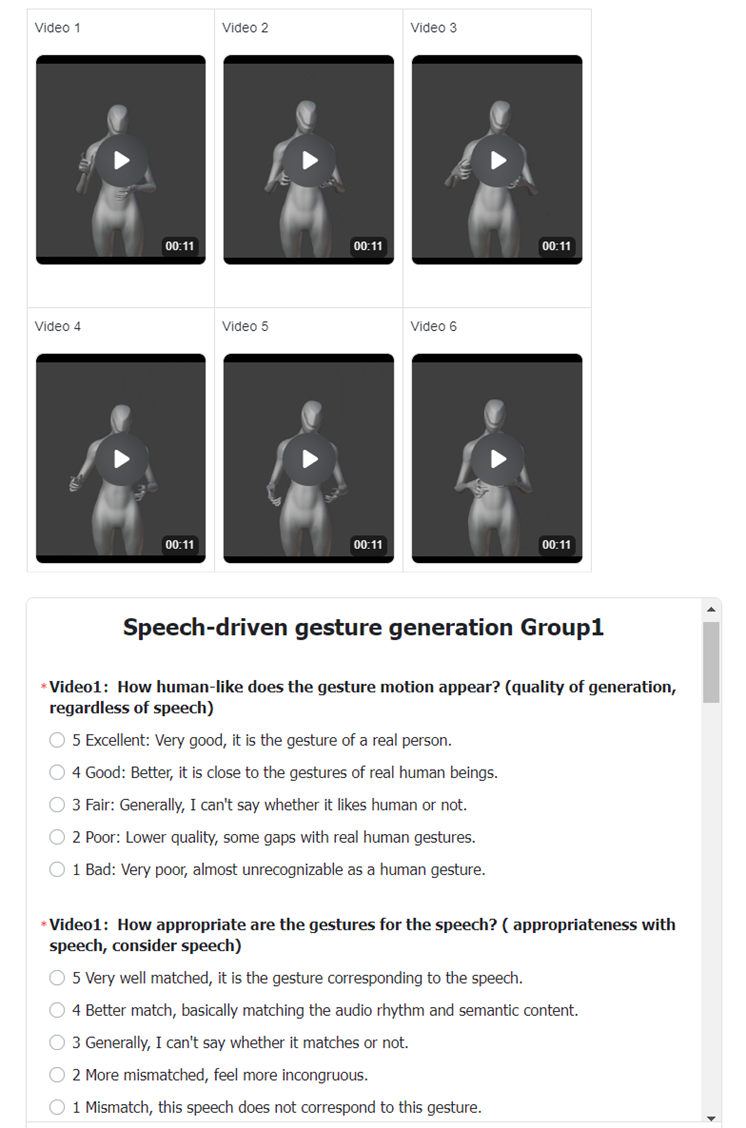} 
    \caption{Screenshot of the parallel rating interface from the user study for ablation studies.}
  \label{fig:ablation studies}
\end{figure}

\begin{figure}[t]
  \centering
  \begin{subfigure}{\linewidth}
  \centering
    \includegraphics[width=\textwidth,height=0.4\textwidth]{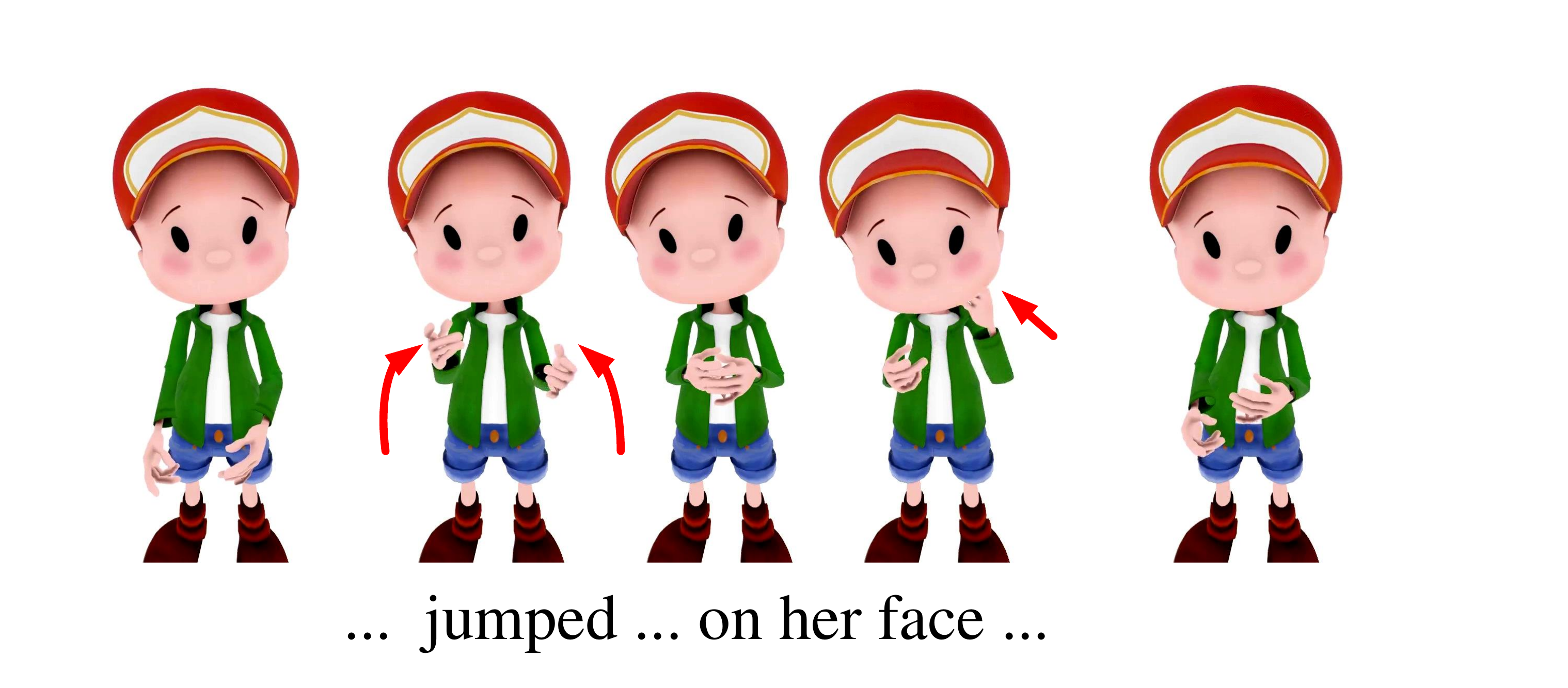}
    \caption{The character makes metaphoric gestures when saying ``jumped'' and deictic gestures for ``face''.}
    \label{fig:aa}
  \end{subfigure}
  \hfill
  \begin{subfigure}{\linewidth}
  \centering
  \includegraphics[width=\textwidth,height=0.4\textwidth]{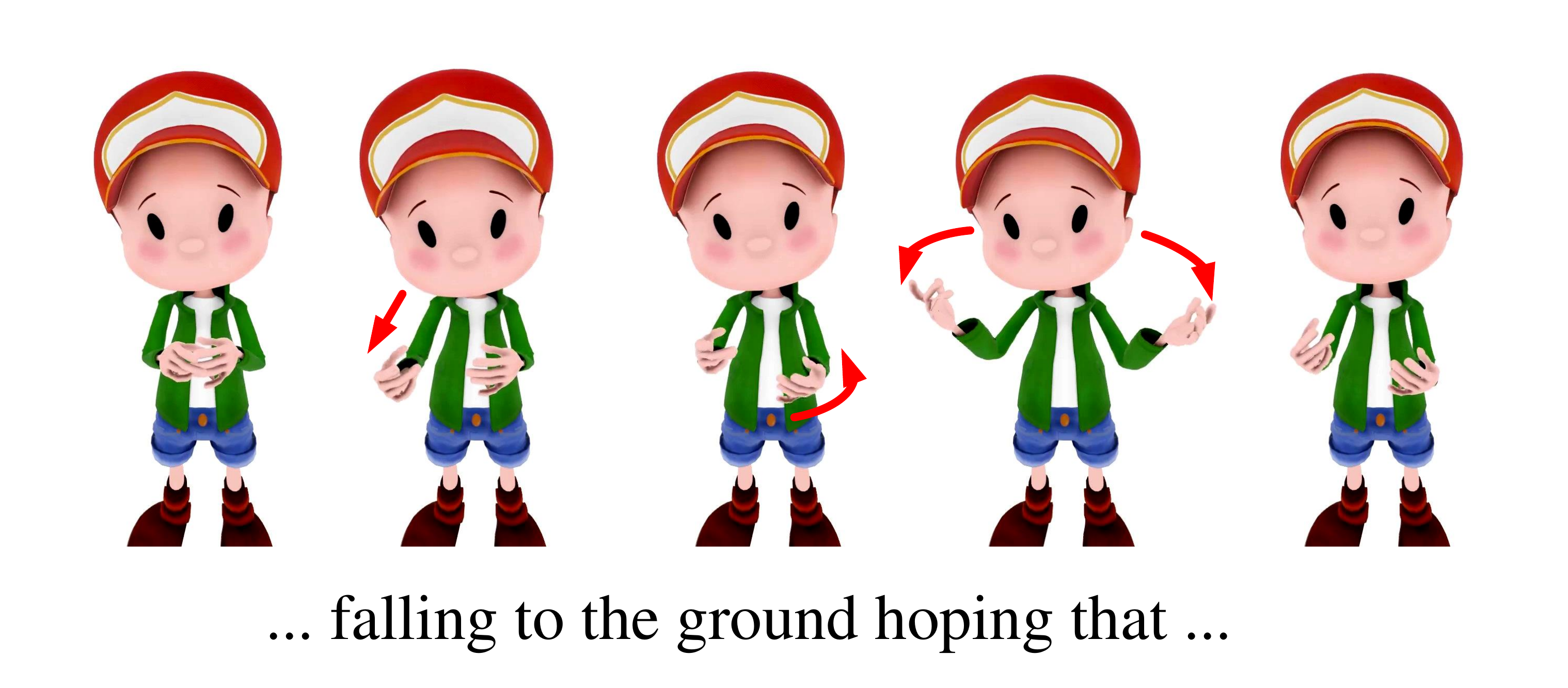}
  \caption{The character makes beat gestures when saying ``falling”, ``ground'' and ``hoping''.}
  \label{fig:b}
  \end{subfigure}
  \caption{Sample results of co-speech gesture generation from our method. Motion history images for some parts are depicted along with the speech text.}
  \label{fig:6}
\end{figure}

The experiment is conducted with 23 participants with good English proficiency to evaluate the human-likeness and appropriateness.
We use two avatar characters to test the robustness of the results, both of them are publicly accessible.
During the evaluation, we prompted the participants to ignore the finger movements and lower body movements, as well as to ignore the problems in skeletal rigging and to pay attention to the upper body gestures.
For human-likeness, it is mainly to evaluate whether the motion of the avatar looks like the motion of a real human.
In terms of appropriateness, it is the evaluation of whether the motion of the avatar is appropriate for the given speech.
A screenshot of the evaluation interface used for comparison with existing methods is presented in Figure \ref{fig:comparison with existing methods}.
An example of the evaluation interface for ablation studies can be seen in Figure \ref{fig:ablation studies}.
Participants reported that the gestures generated by our framework contain many semantic and rhythmically related gestures, as shown in the figure \ref{fig:6}.
Please refer to our supplementary video for comparisons with the baseline model and ablation studies.

\section{Controllability}
\label{Supplementary-Controllability}
\begin{figure}[t]
  \centering
  \begin{subfigure}{\linewidth}
  \centering
    \includegraphics[width=\textwidth,height=0.3\textwidth]{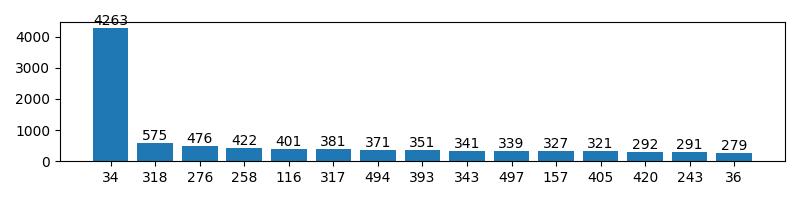}        
    \caption{The horizontal axis indicates the 15 codes with the highest frequency, and the vertical axis indicates the counts.}
    \label{fig:a}
  \end{subfigure}
  \hfill
  \begin{subfigure}{\linewidth}
  \centering
  \includegraphics[width=0.98\textwidth,height=0.35\textwidth, ]{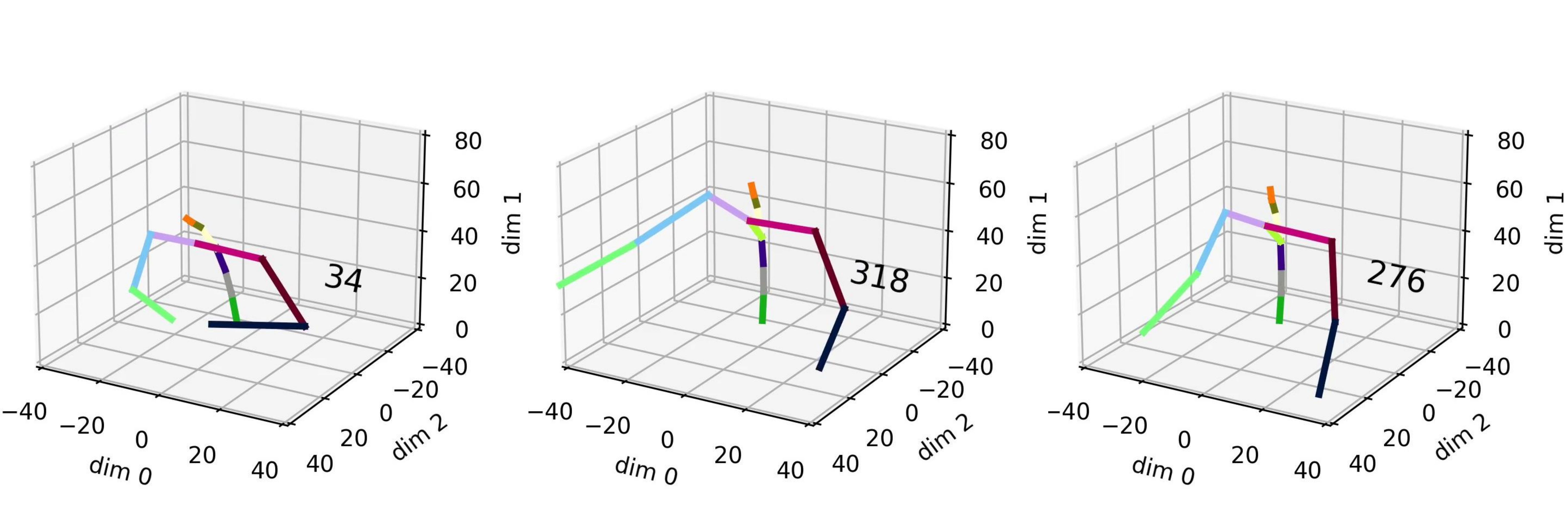}
  \caption{3D joints visualization of the first three codes.}
  \label{fig:bbb}
  \end{subfigure}
  \caption{The histogram of the first 15 code frequencies of speaker ``wayne'' and 3D joints visualization results of the first three codes.}
  \label{fig:Control}
\end{figure}

For the speaker ``wayne", the histogram of the first 15 code frequencies is shown in the Figure \ref{fig:Control}.
It can be seen that the most frequent code is `34', which can be considered to represent the average gesture, that is, the gesture without speech and in silence.
We visualized the three most frequent codes: `34', `318' and `276', and we can find that `318' is a code with a preference for right-handedness. We chose a very typical motion clip using the right-handedness (72s to 76s of gesture ``1\_wayne\_0\_87\_94''), a 4s video with a total of 30 codes at 60FPS and 8 codebook sampling rates, of which there are twelve `318' codes. We use a code with a preference for left-handedness instead of `318' (e.g. `260'), and the results are shown in our supplementary video.

{\small
\bibliographystyleA{ieee_fullname}
\bibliographyA{egbib.bib}
}

\end{document}